\documentclass[reprint, amsmath, amssymb, aps, prb, superscriptaddress]{revtex4-2}
\usepackage{tabularx}
\usepackage{graphicx}
\usepackage{dcolumn}
\usepackage{bm, color}
\usepackage{amsmath}
\usepackage{amsfonts}
\usepackage{amssymb}
\usepackage{amsmath}
\usepackage{multirow}
\providecommand{\U}[1]{\protect\rule{.1in}{.1in}}

\newcommand{\vv}[1]{\boldsymbol #1}

\usepackage{hyperref}
\hypersetup{
	colorlinks = true,
}
\usepackage{physics}

\begin{document}

\title{Majorana multipole response with magnetic point group symmetry}

\author{Yuki Yamazaki}
\affiliation{Condensed Matter Theory Laboratory, RIKEN CPR, Wako, Saitama 351-0198, Japan}
\author{Shingo Kobayashi}
\affiliation{RIKEN Center for Emergent Matter Science, Wako, Saitama 351-0198, Japan}
\date{\today}

\date{\today}

\begin{abstract}
Majorana fermions (MFs) in a topological superconductor exhibit anisotropic electromagnetic responses, called Majorana multipole responses, when MFs are degenerate under time-reversal and crystalline symmetries. In time-reversal symmetric systems, the Majorana multipole response relates to Cooper pair symmetry in the underlying superconducting material, which provides a way to identify pairing symmetries through surface-spin-sensitive measurements.
Here, we extend the concept of Majorana multipole response to systems with magnetic point group symmetry that break time-reversal symmetry and clarify how the response of MFs includes information about underlying superconductors. From a topological classification of symmetry-protected MFs and an effective surface theory, we classify possible magnetic and electric responses for MFs, which manifests a direct connection to Cooper pair symmetry for a symmetry-enforced pair of MFs. Additionally, we find several time-reversal-even higher-order multipole responses, such as the quadrupole response, which are forbidden in time-reversal symmetric systems, whereby indicating breaking of time-reversal symmetry.
The theory is applied to the odd-parity chiral superconductor UTe$_2$ and the ferromagnetic superconductor UCoGe, demonstrating the appearance of a magnetic quadrupole response on a surface.
\end{abstract}

\maketitle

\makeatletter
\def\ext@table{}
\makeatother
\makeatletter
\def\ext@figure{}
\makeatother

\section{Introduction}\label{introduction}
The exploration of unconventional superconductors has been reinvigorated by the study of topological phases of matter. Unconventional superconductors with a topological invariant, called topological superconductors (TSCs), can host Majorana fermions (MFs), which are zero energy self-conjugated fermionic excitations, on a boundary~\cite{Qi-Zhang2011,Tanaka12,Chiu16,SatoFujimoto16,Sato17}. Several types of MFs in TSCs have been proposed theoretically based on their degeneracy and the symmetries that MFs respect. For example, a spinless MF in spin-polarized (spinless) $p$-wave superconductors is non-degenerate and solely protected by particle-hole symmetry (PHS)~\cite{Read10267,Kitaev01}, while a spinful MF in time-reversal invariant TSCs is doubly degenerate and protected by both PHS and time-reversal symmetry (TRS)~\cite{Qi2009, Arbel2019}.

The non-degenerate MF is immune to local noise as long as the superconducting gap is maintained, rendering it suitable for application as a topological qubit in fault-tolerant quantum computation~\cite{Nayak2008}. In contrast, degenerate MFs possess internal degrees of freedom, such as spin, and may react to external fields. In fact, spinful MFs can couple to an applied magnetic field and exhibit a completely anisotropic magnetic response~\cite{Sato094504, shindou10, Chung235301, Nagato123603,  Mizushima12,Dumitrescu245438}, which arises from the self-conjugate property of MFs and TRS. 

In addition to TRS, systems often have other symmetries specific to their crystal structure, such as mirror-reflection and rotation symmetries. The presence of these crystalline symmetries enriches TSC phases, leading to crystalline-symmetry-protected MFs on a surface~\cite{Ueno2013,chiu-yao-ryu, morimoto-furusaki, shiozaki14, Benalcazar224503}. These MFs exhibit various electromagnetic responses under crystalline symmetries~\cite{shiozaki14, xiong17, Mizushima2017, kobayashi, yamazaki043703, yamazaki094508, kobayashi224504, yamazaki073701, yamazakiL060505, Kobayashi2024}, referred to as Majorana multipole responses~\footnote{The term ”magnetic multipole” we employed differs from that commonly used in the literature. In this paper, it refers to the magnetic field angle dependence of the energy gap of MFs.}.

An essential aspect of Majorana multipole responses is their relationship to pairing symmetries in underlying TSCs. The stability of crystalline-symmetry-protected MFs is ensured by a crystalline-symmetry-protected topological invariant via the bulk-boundary correspondence,  implying that the degenerate MFs contain information about bulk superconductors, including pairing symmetries. It has been shown in the previous study~\cite{kobayashi224504} that the magnetic structures of a spinful MF share the same irreducible representations (IRs) with the Cooper pairs under crystalline symmetries. This correspondence suggests that Cooper pair symmetry can be measured through surface spin-sensitive measurements such as spin susceptibility~\cite{Chung235301, Nagato123603, Mizushima12, Mizushima2014, ohashi2024}. This novel application of MFs is significant because determining pairing symmetries is a central issues in the study of unconventional superconductors. Many pairing symmetries in unconventional superconductors remain unidentified except in a few cases, such as the high-$T_{\rm c}$ cuprate.

The studies of the Majorana multipole responses have mainly focused on time-reversal-invariant TSCs because spinful MFs naturally appear due to TRS. However, they have not explored much about time-reversal-symmetry-breaking (TRSB) TSCs. Recently, the topological classification has been extended to the systems with magnetic space group symmetry, predicting various crystalline-symmetry-protected topological phases in TRSB TSCs~\cite{Zou169, Shiozaki2022, Zhang10225, Shiozaki01827}. Additionally, recent experiments on uranium-based superconductors have suggested the potential existence of TRSB TSCs~\cite{Huy067006, Hattori066403,Aoki061011,Jiao523,Bae2644,Ishihara2966}.  This motivates the study of how crystalline-symmetry-protected MFs without TRS react to external fields and to establish the relationship between Majorana multipole responses and pairing symmetries in TRSB TSCs.

In this paper, we extend the theory of Majorana multipole responses to TRSB TSCs with magnetic point group symmetry. First, we classify all possible degenerate MFs under magnetic point group symmetry from the viewpoint of group theory and topological invariants that are relevant to MFs at a high-symmetry point. We show that types of degeneracy are categorized into (a) accidental degeneracy and (b) symmetry-enforced degeneracy. Whereas MFs in type (b) share properties with spinful MFs, those in type (a) do not. Thus, MFs in type (a) occur only in TRSB TSCs. We develop a theory of electromagnetic coupling for degenerate MFs based on an effective surface theory and establish classification tables of electromagnetic responses for two MFs. The classification tables indicate that MFs in type (b) maintain a one-to-one correspondence with pairing symmetries under crystalline symmetries, while those in type (a) allow time-reversal-even higher-order multipole responses that are forbidden in the case of a spinful MF, thereby signaling the breaking of TRS.

We consider two models of TRSB TSCs: a model of odd-parity chiral superconductivity in UTe$_2$ and ferromagnetic superconductivity in UCoGe. These models have point nodes in the superconducting gap, which lead to chiral Majorana edge modes on a surface. By carefully examining the surface energy spectra and topological invariants of tight-binding models with magnetic point group symmetry, we show that the chiral Majorana edge modes are doubly degenerate at a high-symmetry point and exhibit a magnetic quadrupole response.

This paper is organized as follows. 
Section \ref{sec:pre} introduces magnetic point groups and our notations of symmetry operations. Section~\ref{sec:MFs} discusses the topological classification of MFs at a high-symmetry point under magnetic point group symmetry and their degeneracy. Section~\ref{sec:response} develops an effective surface theory for degenerate MFs and classifies possible electromagnetic responses of two MFs protected by magnetic point group symmetry. Section~\ref{sec:app} applies this theory to the odd-parity chiral superconducting state in UTe$_2$ and the ferromagnetic superconducting state in UCoGe.
Section~\ref{sec:conclusion} presents our conclusions. The appendices include the formulation of the Wigner's test in Sec.~\ref{app:wigner}, a detailed explanation of representations of MFs in Sec.~\ref{app:representation}, and a toy model of TRSB TSCs with the magnetic tetrahexacontapole response, the highest order multipole response in our classification, in Sec.~\ref{sec:tetrahexacontapole}. 
 
\section{Summary of key results}
This paper develops a theory of Majorana multipole responses in systems with broken TRS. The formulas for determining the electromagnetic structures of degenerate MFs are provided in Eqs.~(\ref{eq:chi0}) and (\ref{eq:chi_Tg}). These formulas allow us to determine the electromagnetic responses based on pairing symmetries and topological invariants in the bulk.
Before going into the main discussion, we present a brief summary of the main results of the electromagnetic responses in TRSB TSCs.

{\it Electromagnetic responses of two MFs.}  
We focus on magnetic point group symmetry-protected MFs at high-symmetry points in the surface Brillouin zone (BZ) of three-dimensional (3D) TRSB TSCs. The electromagnetic structures of two MFs in systems with type (I) and (III) Shubnikov groups are classified in Tables~\ref{classification_type1} and~\ref{classification_type3}. The main findings are summarized as follows.

\begin{itemize}
\item[(i)]  Both magnetic and electric structures are allowed for two MFs in TRSB TSCs, which contrasts sharply with time-reversal invariant TSCs that allow only magnetic structures. Therefore, a time-reversal-even response is unique to TRSB TSCs and indicates the breaking of TRS.
\item[(ii)] The electromagnetic structures share the same IR with the Cooper pair under crystalline symmetry when the surface symmetries are $mm2$, $3m$, $4mm$, and $6mm$ for type-(I) groups and $4'$, $6'$, $4'm'm$, and $6'mm'$ for type-(III) groups. The notation of magnetic point groups follows the Hermann-Mauguin notation.
\item[(iii)] For type (I) groups, two MFs exhibit magnetic dipole, quadrupole, and octupole responses, with response functions of a magnetic field being of the order of $O(|\bm{B}|)$, $O(|\bm{B}|^2)$, and $O(|\bm{B}|^3)$, respectively. The magnetic quadrupole response is time-reversal-even and occurs in four-fold symmetric systems.
\item[(iv)] For type (III) groups, two MFs exhibit magnetic dipole, quadrupole, octupole, hexadecapole, and tetrahexacontapole responses, with response functions being of the order of $O(|\bm{B}|)$, $O(|\bm{B}|^2)$, $O(|\bm{B}|^3)$, $O(|\bm{B}|^4)$, and $O(|\bm{B}|^6)$, respectively. The magnetic quadrupole, hexadecapole, and tetrahexacontapole responses are time-reversal-even, thus allowed only in TRSB TSCs. These time-reversal-even higher-order multipole responses occur when two MFs are characterized by multiple 1D winding numbers. 
\end{itemize}

{\it Application to UTe$_2$ and UCoGe.}
We apply our theory to a model of the odd-parity chiral superconductor UTe$_2$ and the ferromagnetic superconductor UCoGe. They have crystal-symmetry-protected MFs at a high-symmetry point in a surface BZ, and the surface symmetries that protect them are $m'2'm$ and $m'$, respectively. These models display the magnetic quadrupole response through different mechanisms. In UTe$_2$, the crystal symmetry forbids the linear order terms, so the response appears in the leading order. In UCoGe, the leading order response is the magnetic dipole, but the magnetization enhances the coupling constant of the second order terms, making the magnetic quadrupole response dominant.   

\section{Preliminary}
\label{sec:pre}
\subsection{Magnetic point group}
Here, we summarize symmetries considered in this paper. 
We consider a 3D TRSB superconductor with magnetic point group symmetry, where breaking of TRS is attributed to the coexistence of magnetic order or TRSB pair potentials. 
Generally, magnetic point groups, which consist of 122 groups, are categorized into three types of Shubnikov groups as follows:
\begin{enumerate}
    \item[(I)] ordinary point groups (32),
    \item[(II)] grey point groups (32),
    \item[(III)] black-and-white point groups (58),
\end{enumerate}
where the number in the parentheses represents the number of groups. The magnetic point groups in types (I) and (II) correspond to the ordinary point groups without and with TRS, whereas these in type (III) include the combined operations of point groups and time-reversal operation. When $G$ represents a point group in type (I), the magnetic point groups in type (II) and (III) are represented by, using the time-reversal operator $T$,
\begin{align}
&M^{\text{(II)}} = G + T G \\
&M^{\text{(III)}} = H + T (G-H), 
\end{align}
where, in type (III), $H$ is a unitary subgroup of $G$, while $T(G-H)$ represents an antiunitary subgroup constructed from an element of $G-H$ and $T$.

We consider the 2D magnetic point groups that preserve a surface. The corresponding point groups are 31 magnetic point groups, which consist of 10 type-(I) groups, 10 type-(II) groups, and 11 type-(III) groups. Types-(I) and-(II) groups are given by
\begin{align}
M_0^{\text{(I)}} &\equiv G_0 \\
M_0^{\text{(II)}} &\equiv G_0 +T G_0,
\end{align}
with $G_0 \in 1, 2, 3, 4, 6, m, mm2, 3m, 4mm, 6mm$. 
Here, we have adopted the Hermann-Mauguin notation of magnetic point groups, e.g., $2=\{e,2_z\}$ and $mm2 =\{e,2_z,m_{yz},m_{zx}\}$, and $\{\cdots\}$ describes the generators: $e$ is a unit element, and $n_i$ and $m_{ij}$ represent a $n$-fold rotation operation about the $i$ axis and a mirror-reflection operation in terms of the $ij$ plane, respectively. Type (III) groups are given by extending $G_0$ to $H_0 + T (G_0 -H_0)$ for a given subgroup $H_0 \subset G_0$. We obtain
\begin{align}
     &2', 4',6', m', 3m', m'm2', m'm'2, \notag \\
    &4'm'm, 4m'm', 6'mm',6m'm' \text{ in } M_0^{\text{(III)}}, 
\end{align}
 where the prime implies an element of $T(G_0-H_0)$, e.g., $2'=\{e;T2_z\}$ and $m'm2'=\{e,m_{zx};T2_z,Tm_{yz}\}$. Here, we used the notation $M^{\text{(III)}}=\{H;T(G-H)\}$. $H_0$ in $M_0^{\text{(III)}}$ is shown in Table~\ref{Tab:top_inv_type3}.
In the following, we focus on superconducting states with broken TRS. The corresponding magnetic point groups are type (I) and (III) groups.   
We assume that MFs are on the $(001)$ surface unless otherwise specified.  

\subsection{Symmetry operator in superconducting states}
We start with the Bogoliubov-de Gennes (BdG) Hamiltonian, which is given by
\begin{align}
H_{\text{BdG}} &= \frac{1}{2}\sum_{\bm{k}} \bm{c}^{\dagger}_{\bm{k}} H(\vv k) \bm{c}_{\bm{k}}, \notag \\
H(\vv k) &=  \left[ 
\begin{array}{cc}
  \epsilon(\vv k) & \Delta(\vv k) 
  \\
  \Delta^{\dagger}(\vv k) & - \epsilon^{\rm T}(-\vv k)
\end{array} 
\right],
\label{BdG22}
\end{align}
where $\bm{c}_{\bm{k}} = [c_{\bm{k}},c^{\dagger}_{-\bm{k}}]^{\text{T}}$, and the indices for the spin, orbital, and sublattice degrees of freedom are implicit. $\epsilon(\vv k )$ and $\Delta(\vv k)$ are the normal-state Hamiltonian and gap function, where the gap function satisfies $\Delta^{\rm T}(\vv k) =-\Delta(-\vv k)$ due to the Fermi-Dirac statistics. The BdG Hamiltonian (\ref{BdG22}) is invariant under PHS,
\begin{align}
 C H(\vv{k}) C^{-1} = -H(-\vv{k}), \ \ C=\tau_x K
\end{align}
where $\tau_i$ $(i=x,y,z)$ are the Pauli matrices in Nambu space, and $K$ is the complex conjugate operation.

We consider a superconducting state with magnetic point group symmetry, with an element of $M_0^{\text{(I)}}$ and $M_0^{\text{(III)}}$ acting on Eq.~(\ref{BdG22}) as follows. First, we consider type (I) groups, which consist of unitary group operations. For a given $g \in G_0$, the operations are defined by
\begin{align}
    &D(g) \epsilon(\vv k) D^{\dagger}(g) = \epsilon(g \vv k),  \label{eq:normalstateg} \\
    &D(g) \Delta(\vv k) D^{\rm T}(g) = \eta_g \Delta(g \vv k), \label{eq:gapfunctiong}
\end{align}
where $g\vv{k}$ means the $O(3)$ transformation of $\vv{k}$ in terms of $g$, say, $2_z {\vv k} = (-k_x,-k_y,k_z)$, and $D(g)$ is a spinful unitary representation of $g$. We assume that the gap function belongs to a one-dimensional (1D) IR of $M_0^{\text{(I)}}$. The factor $\eta_g \in U(1)$ encodes the information about IRs that the gap function belongs to. For instance, when $g \in 2 = \{e,2_z\}$, $\eta_{2_z} = +1 (-1)$ for the $A$ ($B$) IR of $2$.  In the Nambu space, Eqs.~(\ref{eq:normalstateg}) and (\ref{eq:gapfunctiong}) are represented as
\begin{align}
    &\tilde{D}(g) H(\vv k) \tilde{D}^{\dagger}(g) = H(g\vv k), \notag \\
    &\tilde{D}(g) \equiv  \left[ \begin{array}{cc} D(g) & 0 \\ 0 & \eta_g D^{\ast}(g) \end{array} \right], \label{eq:nambug} 
\end{align}
where $C$ and $\tilde{D}(g)$ satisfy 
\begin{align}
    \tilde{D}(g)C = \eta_g C \tilde{D}(g), \label{eq:commu_CD}
\end{align}
implying that the information about IRs is included in the commutation relation.

On the other hand, type-(III) groups consist of unitary and antiunitary group operations. An element in $H_0$ is represented by a unitary group operation and acts as shown in Eq.~(\ref{eq:nambug}), whereas an element in $T(G_0-H_0)$ is represented by an antiunitary group operation such that, for a given $h \in G_0 -H_0$,  
\begin{align}
    &D(Th) \epsilon^{\ast}(\vv k) D^{\dagger}(Th) = \epsilon(-h \vv k),  \label{eq:normalstatem} \\
    &D(Th) \Delta^{\ast}(\vv k) D^{\rm T}(Th) = \eta_{Th} \Delta(-h \vv k), \label{eq:gapfunctionm}
\end{align}
where $D(Th)$ is a corepresentation of $M_0^{\text{(III)}}$. We assume that the gap function belongs to a 1D IR~\footnote{Our theory extends to pairing symmetries in higher dimensional IRs. For example, consider the $E$ IR of $4mm$, where $p_x$ and $p_y$ states are degenerate. The time-reversal symmetry breaking order parameters with lower free energies can be described by $p_x \pm ip_y$ states, each of which breaks the $4mm$ symmetry. Focusing on the $p_x + ip_y$ state, it remains invariant under the $4 \subset 4mm$ and belongs to the $^2E$ IR.}, i.e., the transformation of the gap function under $M_0^{\text{(III)}}$ is characterized by $\eta_{Th} \in U(1)$. We note that although $\eta_{Th}$ takes an arbitrary value due to the gauge degrees of freedom of the gap function, the relationship between $\eta_{Th}$ and  $\eta_{Th'}$ ($h \neq h'$) is given by $\eta_{g}$. For instance, if the gauge is fixed such that $\eta_{Th}=1$ for a given $h \in G_0-H_0$, the other factors are given by $\eta_{Th'}=\eta_g^{\ast}$ with $h' = h g$. Thus, the gap functions are classified by IRs of $H_0$.
In the Nambu space, Eqs.~(\ref{eq:normalstatem}) and (\ref{eq:gapfunctionm}) are represented as
\begin{align}
    &\tilde{D}(Th) H^{\ast}(\vv k) \tilde{D}^{\dagger}(Th) = H(-h\vv k), \notag \\
    &\tilde{D}(Th) \equiv  \left[ \begin{array}{cc} D(Th) & 0 \\ 0 & \eta_{Th} D^{\ast}(Th) \end{array} \right]. \label{eq:nambugTh} 
\end{align}
Here, $\tilde{D}(Th)$ also satisfies the commutation relation,
\begin{align}
    \tilde{D}(Th)C = \eta_{Th} C \tilde{D}(Th). \label{eq:commu_CDT}
\end{align}

\section{Majorana fermions protected by magnetic point groups}
\label{sec:MFs}
\subsection{1D topological classification}
According to the ten-fold way topological classification~\cite{schnyder,kitaev,Schnyder09,ryu}, a TRSB superconductor is classified into class D within the Altland-Zirnbauer symmetry class~\cite{Altland1142}. In this class, no three-dimensional (3D) topological invariant exists. However, when magnetic point group symmetry is considered, the topological classification expands, predicting a wide range of crystalline topological phases in 3D. These phases fall into two categories: fully-gapped and gapless. In fully-gapped topological phases, isolated MFs appear at high-symmetry points or lines in a surface BZ with magnetic point group symmetry. The MFs are characterized by crystalline-symmetry-protected topological invariants defined in low-dimensional subspaces. On the other hand, gapless topological phases host point or line nodes in the bulk BZ, and a flat band state of MFs associated with these nodes appears in a surface BZ. Gapless topological phases are characterized by the Chern number and crystalline-symmetry-protected topological invariants in a low dimension, where the Chern number is nonzero when the point nodes appear, accompanying them. 
Here, we examine MFs at a high-symmetry point in a surface BZ with magnetic point group symmetry. These MFs are supported by 1D crystalline symmetry-protected topological invariants in both fully-gapped and gapless topological phases.
 
The classification of 1D topological invariants under magnetic point groups $M_0^{\text{(I)}}$ and $M_0^{\text{(III)}}$ is systematically performed using the Wigner's test. The details of the calculation are relegated to Appendix~\ref{app:wigner}. Possible 1D topological invariants are summarized in Tables~\ref{Tab:top_inv_type1} and \ref{Tab:top_inv_type3}, where $\nu[g]_{j}$ and $W[g']^g_{j}$ are, respectively, crystalline-symmetry-protected $\mathbb{Z}_2$ and $\mathbb{Z}$ topological invariants defined in an eigenspace of $\tilde{D}(g)$ with eigenvalue $e^{-i\pi(2j-1)/|g|}$ ($j=1,\cdots, |g|$), where $|g|$ is the order of $g$, e.g., $|n_z| = n$. 

In the following, we discuss examples of the $\mathbb{Z}_2$ and $\mathbb{Z}$ topological invariants. First, consider the case with $M_{0}^{\text{(I)}} = 2=\{e,2_z\}$ and the $B$ IR ($\eta_{2_z} = -1$); namely, the unitary representation $\tilde{D}(2_z)$ satisfies $\tilde{D}^2(2_z)=-1$ and $\{C,\tilde{D}(2_z)\}=0$. In this case, we can define a $\mathbb{Z}_2$ topological invariant as follows. On a $2_z$-symmetry-invariant 1D subspace specified by $k$, $\tilde{D}(2_z)$ commutes with $H(k)$. Hence, $H(k)$ can be block-diagonalized as $H_1(k) \oplus H_2(k)$ in the basis diagonalizing $\tilde{D}(2_z)$, where the subscript represents the eigenvalues of $\tilde{D}(2_z)$. Let $|n,k,j \rangle$ ($j=1,2$) be an eigenstate of $H_{j} (k)$ satisfying $H_{j} (k)|n,k,j \rangle =E_n |n,k,j \rangle$ ($E_n \in \mathbb{R}$) and $\tilde{D}(2_z)|n,k,j \rangle =e^{-i (2 j -1)\pi/2} |n,k,j \rangle$.  Since $\{C,\tilde{D}(2_z)\}=0$,
\begin{align}
     \tilde{D}(2_z) C|n,k,j \rangle =   e^{-i (2 j -1)\pi/2} C |n,k,j \rangle,
\end{align}
so the PHS is preserved within $H_{j}$. Thus, a 1D $\mathbb{Z}_2$ invariant $\nu[2_z]_{j}$ is defined in each eigenspace,
\begin{align}
&\nu[2_z]_{j} \equiv \int_{-\pi}^\pi \frac{dk}{2\pi} a_{j}(k) \text{ mod } 2, \\
&a_{j}(k) \equiv -i\sum_{E_n<0} \langle n,k,j|\frac{\partial}{\partial k}|n,k,j \rangle,
\label{Z2-invariant}
\end{align}
where $a_{j}(k)$ is the Berry connection in $H_j(k)$, and $\nu[2_z]_{j=1,2}$ take $0,1$ due to the PHS. 
In general, when the BdG Hamiltonian commutes with $\tilde{D}(g)$, it is block-diagonalized as $H_1(k) \oplus \cdots \oplus H_{|g|}(k) $ in the eigenspaces of $\tilde{D}(g)$. The eigenstate satisfies $\tilde{D}(g)|n,k,j \rangle = e^{-i\pi(2j-1)/|g|}|n,k,j \rangle$. A $\mathbb{Z}_2$ topological invariant $\nu[g]_j$ is defined in $H_j(k) $ if the eigenstate satisfies  
\begin{align}
   \tilde{D}(g) C|n,k,j \rangle =   e^{-i (2 j -1)\pi/|g|} C |n,k,j \rangle.
\end{align}
 Under type-(I) groups, we only have the $\mathbb{Z}_2$ topological invariants because no antiunitary operator exists. Note that $\nu[g]_j$ and $\nu[g]_{j'}$ ($j' \neq j$) are related to each other when the degeneracy of MFs is enforced by $M_0^{\rm (I)}$.

An example of the $\mathbb{Z}$ topological invariant is the case with $M_0^{\text{(III)}} = 2'=\{e;T2_z\}$ and the $A$ IR. The combination of $\tilde{D}(T2_z)$ and $C$ gives a chiral operator,
\begin{align}
   \{\Gamma(T2_z), H(k)\}=0, \ \ \Gamma(T2_z) \equiv e^{i \phi} \tilde{D}(T2_z) \tau_x,
\end{align}
on a symmetric 1D subspace. Here $\phi$ is determined to be  $\Gamma^2(T2_z) =1$. Thus, we can define a 1D winding number using the chiral operator,
\begin{align}
&W[2_z] \equiv \frac{i}{4\pi} \int_{-\pi}^\pi dk \tr\left\{\Gamma(T2_z) H^{-1}(k) \pdv{H(k)}{k}\right\},
\label{magnetic-winding-number}
\end{align}
where $W[2_z] \in \mathbb{Z}$, and we omit the suffix since $H_0 = 1$. When $H_0 \neq 1$, a 1D winding number is defined as follows. The BdG Hamiltonian is block-diagonalized into $H_1(k) \oplus \cdots \oplus H_{|g|}(k) $ in the eigenspaces of $\tilde{D}(g)$. 
If a chiral operator $\Gamma(Th) \equiv e^{i \phi}\tilde{D}(Th)\tau_x$ commutes with $\tilde{D}(g)$, it is also block-diagonalized into $ \Gamma_1(Th) \oplus \cdots \oplus \Gamma_{|g|}(Th)$ and $\{\Gamma_{j}(Th),H_j (k)\}=0$. Then, a 1D winding number in $H_j(k)$ is defined as
\begin{align}
&W[h]^g_{j} \equiv \frac{i}{4\pi} \int_{-\pi}^\pi dk \tr\left\{\Gamma_j(Th) H_j^{-1}(k) \pdv{H_j(k)}{k}\right\},
\label{magnetic-winding-number_g}
\end{align}
where $\Gamma_j(Th)$ satisfies $\Gamma_j^2(Th)=e^{-2i \phi_j}$ for $\phi_j \in \mathbb{R}$, and $e^{i \phi_j}W[h]^g_{j} \in \mathbb{Z}$.
The explicit definition of $W[4_z]^{2_z}_{j}$ and $W[6_z]^{3_z}_{j}$ are shown in Appendix~\ref{app:representation}. Note that $W[h]^g_{j}$ and $W[h]^g_{j'}$ ($j' \neq j$) are related to each other when the degeneracy of MFs is enforced by $M_0^{\text{(III)}}$; e.g., for $M_0^{\text{(III)}} = 4'$, $|W[4_z]^{2_z}_{1}|=|W[4_z]^{2_z}_{2}|$.

\begin{table}
	\caption{The 1D topological invariants and degeneracy of MFs for type (I) groups $M_0^{\text{(I)} }$, where each column shows the magnetic point group, 1D IRs of the gap functions, 1D topology, associated topological invariants, and degeneracy of MFs. $\nu[g]_{j}$ represents the $\mathbb{Z}_2$ topological invariant defined in an eigenspace with $e^{-i\pi (2j-1)/|g|}$, where $\nu[g]_{j,j'}$ is the shorthand notation of $\nu[g]_{j}$ and $\nu[g]_{j'}$. The types of degeneracy of MFs are categorized into (a) accidental degeneracy, (b) symmetry-enforced degeneracy, and no degeneracy. In the $A_2$ IR of $3m$, (a) and (b) cannot be distinguished solely by the topological invariant. The Hermann-Mauguin notation of magnetic point groups is adopted, and the labels of IRs are in the Mulliken notation~\cite{Bradley72}. $n_z$ represents $n$-fold rotation about the $z$-axis, and $m_{ij}$ and $m_d$ represent a mirror reflection in terms of the $ij$ and $(110)$ planes.  }
		\begin{tabular}{lccccc}
                   \hline \hline
			$M_0^{\text{(I)}}$ & IR of $\Delta$ & Topo. & Invariants & Degeneracy
			\\
			\hline
			$2$ & $A$ & $0$ &  & 
                \\
                $2$ & $B$ & $\mathbb{Z}_2$ & $ \nu[2_z]_{1,2}$ & (a)
			\\
                $3$ & $A_1$ & $\mathbb{Z}_2$ & $\nu[3_z]_{2}$ & no
			\\
                $3$ & $^1 E$ & $\mathbb{Z}_2$ & $\nu[3_z]_{1}$ & no
			\\
                $3$ & $^2 E$ & $\mathbb{Z}_2$ & $\nu[3_z]_{3}$ & no
			\\
                $4$ & $A$ & $0$ & &
			\\   
                $4$ & $B$ & $0$ & &
			\\   
                $4$ & $^1 E$ & $\mathbb{Z}_2$ & $\nu[4_z]_{1,3}$ & (a)
			\\   
                $4$ & $^2 E$ & $\mathbb{Z}_2$ & $\nu[4_z]_{2,4}$ & (a)
			\\ 
                $6$ & $A$ & $0$ & &
			\\   
                $6$ & $B$ & $\mathbb{Z}_2$ & $\nu[6_z]_{2,5}$ &  (a)
			\\   
                $6$ & $^1E_1$ & $0$ & &
                \\
                $6$ & $^1 E_2$ & $\mathbb{Z}_2$ & $\nu[6_z]_{1,4}$ & (a)
			\\   
                $6$ & $^2E_1$ & $0$ & &
                \\
                $6$ & $^2 E_2$ & $\mathbb{Z}_2$ & $\nu[6_z]_{3,6}$ & (a)
                \\
		    $m$ & $A'$ & $0$ &  &  &
                \\
               $m$ & $A''$ & $\mathbb{Z}_2$ & $ \nu[m_{zx}]_{1,2}$ & (a)
			\\
   		    $mm2$ & $A_1$ & $0$ &  & 
                \\
              $mm2$ & $A_2$ & $\mathbb{Z}_2$ & $ \nu[m_{zx}]_{1,2},\nu[m_{yz}]_{1,2}$ & (b)
			\\
                $mm2$ & $B_1$ & $\mathbb{Z}_2$ & $ \nu[2_z]_{1,2}, \nu[m_{yz}]_{1,2}$  & (b)
			\\
                $mm2$ & $B_2$ & $\mathbb{Z}_2$ & $ \nu[2_z]_{1,2}, \nu[m_{zx}]_{1,2}$ & (b)
			\\
                $3m$ & $A_1$ &  $0$  &  &
			\\
                $3m$ & $A_2$ & $\mathbb{Z}_2$ &  $ \nu[m_{zx}]_{1,2}$  & (a) and (b)
			\\
                $4mm$ & $A_1$ &  $0$  &  &
			\\
                $4mm$ & $A_2$ & $\mathbb{Z}_2$ & $\nu[m_{zx}]_{1,2}, \nu[m_{d}]_{1,2}$   &  (b) 
			\\
                $4mm$ & $B_1$ &  $0$ &  &
			\\
                $4mm$ & $B_2$ &  $0$  &  &
                \\
                $6mm$ & $A_1$ &  $0$  &  &
			\\
                $6mm$ & $A_2$ & $\mathbb{Z}_2$ & $\nu[m_{zx}]_{1,2}, \nu[m_{yz}]_{1,2}$  &  (b) 
			\\
                $6mm$ & $B_1$ &  $\mathbb{Z}_2$ & $\nu[2_z]_{1,2}, \nu[m_{zx}]_{1,2}$  & (b)
			\\
                $6mm$ & $B_2$ &  $\mathbb{Z}_2$  &  $\nu[2_z]_{1,2}, \nu[m_{yz}]_{1,2}$  & (b)
			\\
              \hline \hline
 \end{tabular}	
\label{Tab:top_inv_type1}
\end{table}

\begin{table}
	\caption{The 1D topological invariants and degeneracy of MFs for type (III) groups $M_0^{\text{(III)}}$, where each column shows the magnetic point groups, unitary groups of $M_0^{\text{(III)}}$, 1D IRs of the gap functions, 1D topology, associated 1D topological invariants, and degeneracy of MFs, where the IRs of $\Delta$ represents the IRs of $H_0$. $ W[h]^g_{j}$ is the $\mathbb{Z}$ topological invariant defined in an eigenspace with $e^{-i\pi (2j-1)/|g|}$. Only a minimal set of topological invariants is shown here. The notations align with those in Table~\ref{Tab:top_inv_type1}.  }
		\begin{tabular}{lcccccc}
                   \hline \hline
			$M_0^{\text{(III)}}$ &$H_0$ & IR of $\Delta$ & Topo. & Invariants & Deg.
			\\
			\hline
			$2'$ &$1$ & $A$ & $\mathbb{Z}$ & $W[2_z]$ & (a)
                \\
                $4'$ & $2$ & $A$ & $\mathbb{Z}$ & $W[4_z]^{2_z}_{1,2}$ & (b)
			\\  
                $4'$ & $2$ & $B$ & $\mathbb{Z}_2$ & $ \nu[2_z]_{1,2}$ & (b)
			\\    
                $6'$ & $3$ & $A_1$ & $\mathbb{Z}$ & $ W[6_z]^{3_z}_{2}$ & (a)
            \\
                $6'$ & $3$ & $A_1$ & $\mathbb{Z}$ & $ W[6_z]^{3_z}_{1,3}$ & (b)
            \\
		      $m'$ & $1$ & $A$ & $\mathbb{Z}$ & $W[m_{zx}]$ & (a)
            \\
   		    $m'm2'$ & $m$ & $A'$ & $0$ &  & 
            \\
                $m'm2'$ & $m$ & $A''$ & $\mathbb{Z}$ & $ W[2_z]^{m_{zx}}_{1,2}, W[m_{yz}]^{m_{zx}}_{1,2}$ & (a)
			\\
   		    $m'm'2$ & $2$ & $A$ & $0$ &  & 
            \\
                $m'm'2$ & $2$ & $B$ & $\mathbb{Z}$ & $ W[m_{yz}]^{2_z}_{1,2}, W[m_{zx}]^{2_z}_{1,2}$ & (a)
			\\
                $3m'$ & $3$ & $A_1$ & $\mathbb{Z}$ & $ W[m_{zx}]^{3_z}_{2}$  & (a)
			\\
                $3m'$ & $3$ & $^1 E$ & $\mathbb{Z}$ & $ W[m_{zx}]^{3_z}_{1}$  & (a)
			\\               
                $3m'$ & $3$ & $^2 E$ & $\mathbb{Z}$ & $ W[m_{zx}]^{3_z}_{3}$  & (a)
			\\
                $4'm'm$ & $mm2$ & $A_1$ &  $0$  &  &
			\\
                $4'm'm$ & $mm2$ & $A_2$ &  $\mathbb{Z}$ & $W[4_z]^{2_z}_{1,2}$  & (b)
			\\
                $4m'm'$ & $4$ & $A$ &  $0$  &  &
			\\
                $4m'm'$ & $4$ & $B$ &  $0$  &  &
			\\
                $4m'm'$ & $4$ & $^1 E$ & $\mathbb{Z}$ & $W[m_{yz}]^{4_z}_{1,3}, W[m_{d}]^{4_z}_{1,3}$ & (a)
			\\
                $4m'm'$ & $4$ & $^2 E$ & $\mathbb{Z}$  & $W[m_{yz}]^{4_z}_{2,4}, W[m_{d}]^{4_z}_{2,4}$ & (a)
			\\
                $6'mm'$ & $3m$ & $A_1$ &  $0$  &  &
			\\
                $6'mm'$ & $3m$ & $A_2$ &  $\mathbb{Z}$  & $W[6_z]^{3_z}_{2}, W[m_{zx}]^{3_z}_{1,3}$ & (a)
            \\
                $6'mm'$ & $3m$ & $A_2$ &  $\mathbb{Z}$  & $W[6_z]^{3_z}_{1,3}$ & (b)
			\\
                $6m'm'$ & $6$ & $A$ &  $0$  &  &
			\\
              $6m'm'$  & $6$ & $B$ &  $\mathbb{Z}$  & $W[m_{zx}]^{6_z}_{2,5}, W[m_{yz}]^{6_z}_{2,5}$ & (a)
			\\
                $6m'm'$  & $6$ & $^1E_1$ &  $0$  &  &
			\\
               $6m'm'$  & $6$ & $^1 E_2$ &   $\mathbb{Z}$  & $W[m_{zx}]^{6_z}_{1,4}, W[m_{yz}]^{6_z}_{1,4}$ & (a)
			\\
                $6m'm'$  & $6$ & $^2E_1$ &  $0$  &  &
			\\
               $6m'm'$  & $6$ & $^2 E_2$ &  $\mathbb{Z}$  & $W[m_{zx}]^{6_z}_{3,6}, W[m_{yz}]^{6_z}_{3,6}$ & (a)
			\\
              \hline \hline
 \end{tabular}	
\label{Tab:top_inv_type3}
\end{table}

\subsection{Degeneracy of Majorana fermions}
\label{sec:degeneracy}
Here, we discuss the degeneracy of MFs at a high-symmetry point in the surface BZ under magnetic point group symmetry. The degeneracy is essential for the Majorana multipole response since a non-degenerate MF cannot respond to any external fields. From the 1D topological classification, there are the $\mathbb{Z}_2$ and $\mathbb{Z}$ topological invariants in TSCs with type (I) and type (III) groups. The bulk-boundary correspondence manifests as the relationship between the number of the topological invariants and the number of MFs. Thus, degeneracy of MFs occurs if the value of the topological invariants are larger than one or more than two topological invariants take a nonzero value. For instance, double degeneracy of MFs occurs when $\nu[g]_{j} = \nu[g]_{j'} =1$,  $|W[h]^g_{j}| = 2$, or $|W[h]^g_{j}| = |W[h]^g_{j'}| = 1$, with $j \neq j'$.  

On the other hand, from the group theoretical viewpoint, there are two types of degeneracy: (a) accidental degeneracy and (b) symmetry-enforced degeneracy. In type (b), the double degeneracy of MFs is enforced by the magnetic point group, while, in type (a), it is not enforced by the magnetic point group, but an accidental degeneracy is allowed. 
The types (a) and (b) are determined from the relationship between the 1D topological invariants under magnetic point group operations, which are summarized in Tables~\ref{Tab:top_inv_type1} and \ref{Tab:top_inv_type3}.

 We note that besides 1D topological invaraints, 2D topological invariants such as the mirror Chern number also appear in 3D superconductors with types (I) and (III) groups. The mirror Chern number is defined on a mirror invariant plane under a mirror reflection symmetry. MFs associated with the mirror Chern number live in the mirror invariant line in the surface BZ, but not restricted to a high-symmetry point, which case is beyond the scope of our paper. 

\section{Majorana multipole response}
\label{sec:response}

To classify the electromagnetic responses of MFs under magnetic point group symmetry, we employ an effective surface theory developed in the previous studies~\cite{kobayashi, kobayashi224504}, which allows us to determine possible electromagnetic responses from the information about representations of MFs subject to crystalline symmetries. Representations of MFs are determined from information about IRs of the gap function and the relevant topological invariants. In the following, we formulate the electromagnetic responses of MFs based on the effective surface theory in low energy and apply the formula to TSCs with type-(I) and (III) groups.

\subsection{Effective surface theory}
We start from a bulk Hamiltonian describing an effective coupling between magnetic field $\bm{B}$ and quantum fields $\bm{c}_{\bm{k}}$. In the low-energy limit, a coupling coefficient is independent of momentum and is given by, in the Numbu space, 
\begin{align}
H_m = \sum_{\bm{k}} f(\bm{B}) \tilde{O}(\bm{k}), \label{eq:m_coupling}
\end{align} 
 with
\begin{align}
\tilde{O}(\bm{k}) &= \frac{1}{2} \bm{c}^{\dagger}_{\bm{k}} \, \tilde{O} \, \bm{c}_{\bm{k}}, \;  \tilde{O}=\left[
\begin{array}{cc}
O & 0 \\
0 & -O^T
\end{array}
\right], \label{eq:general_op}
\end{align}
where $f(\bm{B})$ is a real function of $\bm{B}$, $\bm{c}_{\bm{k}} = [c_{\bm{k}},c^{\dagger}_{-\bm{k}}]^{\text{T}}$, and $O$ is a hermitian operator, i.e., $\{\tilde{O},C\}=0$. We perform the mode expansion in terms of $\bm{c}_{\bm{k}}$ and extract the contribution to $\tilde{O}(\bm{k})$ from MFs: $\bm{c}_{\bm{k}}$ is decomposed into MFs $|u^{(a)}_0 \rangle$ and others:
\begin{align}
\bm{c}_{\bm{k}} = \sum_a \gamma_a|u^{(a)}_0\rangle + \cdots, \label{eq:mode_exp}
\end{align}
where $a$ labels the MFs, and $\gamma_a$ are Majorana operators. We assume that the MFs are located at time-reversal invariant momenta $\bm{k}_0$ in the surface BZ. Substituting Eq.~(\ref{eq:mode_exp}) into Eq.~(\ref{eq:general_op}) and using the relation $\bm{c}^{\dagger}_{\bm{k}} \tau_x = \bm{c}^T_{-\bm{k}} $ and $\bm{k}_0 = - \bm{k}_0$, a quantum operator of MFs becomes~\cite{kobayashi224504} 
\begin{align}
    \tilde{O}_{\text{MF}}(\bm{k}_0) &= \frac{1}{2} \sum_{a,b} \gamma_a \gamma_b \, \text{tr} \left[\tilde{O}|u^{(b)}_0 \rangle \langle Cu^{(a)}_0| \right] \notag  \\
    &=-\frac{1}{8}\sum_{a,b} [\gamma_a,\gamma_b] \text{tr}\left[\tilde{O}\rho^{(ab)}\right], \label{eq:multipole}
\end{align}
where $|C u_0^{(a)} \rangle = \tau_x | u_0^{(a)\, \ast} \rangle$ and  $\rho^{(ab)}$ is defined by
\begin{align}
    \rho^{(ab)} =|u^{(a)}_0\rangle \langle Cu^{(b)}_0|- |u^{(b)}_0\rangle \langle Cu^{(a)}_0|. \label{eq:rho_op}
\end{align}
Equation~(\ref{eq:multipole}) is a key formula to determine electromagnetic responses of MFs, which tells us that the MFs couple to the local operator only when the trace part is nonzero. When $\tilde{O}_{\text{MF}} \neq 0$, an coupling between a magnetic field and MFs is described as
\begin{align}
    H_m = f(\bm{B}) \tilde{O}_{\text{MF}} (\bm{k}_0), \label{eq:magnetic_coupling}
\end{align}
This implies that $f(\bm{B})$ is under the influence of crystalline symmetries since $H_m$ should be invariant under the symmetry operation which acts on Eq.~(\ref{eq:m_coupling}) as 
\begin{subequations}
\label{eq:Btrans_uni}
\begin{align}
&f(\bm{B}) \to f(g \bm{B}), \\
&\tilde{O} \to \tilde{D}(g) \tilde{O} \tilde{D}^{\dagger}(g) 
\end{align}
\end{subequations}
for a unitary operation, and 
\begin{subequations}
\label{eq:Btrans_uni}
\begin{align}
&f(\bm{B}) \to f(-h\bm{B}), \label{eq:response_anti1}\\
&\tilde{O} \to \tilde{D}(Th) \tilde{O}^{\ast} \tilde{D}^{\dagger}(Th)  \label{eq:response_anti1}
\end{align}
\end{subequations}
for an antiunitary operation. Here, $\bm{B}$ flips under the time-reversal operation. 

In a similar manner, we can discuss a coupling between electric perturbations such as strain and MFs. A coupling between strain and MFs is described as
\begin{align}
    H_{\text{s}} = f(\bar{u}) \tilde{O}_{\text{MF}} (\bm{k}_0). \label{eq:strain_coupling}
\end{align}
where $\bar{u}$ is the unsymmetrized strain tensor $\bar{u}_{ij} \equiv \partial_i u_j$ ($i,j = x,y,z$), where $u_i$ is the lattice displacement field, and $\bar{u}$ is the second-rank tensor that transforms as $\bar{u} \to g  \bar{u} g^T$ under the action of $g \in G_0$. Note that the strain tensor $\bar{u}$ does not change under the time-reversal operation, i.e., Eq.~(\ref{eq:strain_coupling}) describes a time-reversal-even response.

\subsection{Representation of $\rho^{(ab)}$}
From the group theoretical argument, $\tilde{O}_{\text{MF}}(\bm{k}_0)$ is nonzero only when $\tilde{O}$ shares the same IR of $\rho^{(ab)}$~\cite{kobayashi224504}. Thus, the representation of $\rho^{(ab)}$ determines how $\tilde{O}$ transforms under the action of crystalline symmetry operations. 

The following formula provides a method to identify the representation of $\rho^{(ab)}$ under magnetic point group operations. We assume that an MF $|u_0^{(a)} \rangle$ is a zero mode of $H(k)$ on a 1D subspace with a magnetic point group. Since the magnetic point group commutes with $H(k)$, $\tilde{D}(g) | u_0^{(a)} \rangle$ is also a zero mode, linking them through a unitary transformation:
\begin{align}
 \tilde{D}(g) | u_0^{(a)} \rangle = \sum_{b}| u_0^{(b)} \rangle [\tilde{D}_{\gamma}(g)]_{ba},
\end{align}
where $\tilde{D}_{\gamma}(g) \equiv \langle u_0^{(a)}| \tilde{D}(g)| u_{0}^{(b)} \rangle $, dubbed short representations~\cite{kobayashi224504}, is an $N \times N$ unitary matrix ($N$ is the number of MFs) and obeys the same multiplication law as $\tilde{D}(g)$.  Using $\tilde{D}_{\gamma}(g)$, the transformation of $\rho^{(ab)}$ under the group operation $g$ is represented as
\begin{align}
    \tilde{D}(g) \rho^{(ab)} \tilde{D}^{\dagger}(g) = \sum_{c,d} \rho^{(cd)} \eta^{*}_g [\tilde{D}_{\gamma}(g)]_{ca} [\tilde{D}_{\gamma}(g)]_{db}, \label{eq:rho_g1}
\end{align}
where we have used Eq.~(\ref{eq:commu_CD}). We emphasize that the representation of $\rho^{(ab)}$ has information about the gap function via $\eta_g$. Since $\rho^{(ab)}$ is a skew symmetric matrix, its symmetric part is zero. Thus, Eq.~(\ref{eq:rho_g1}) is recast into 
\begin{align}
     \tilde{D}(g) \rho^{(ab)} \tilde{D}^{\dagger}(g) \equiv \sum_{c,d} \rho^{(cd)} [\Omega(g)]_{(cd)(ab)},
\end{align}
with
\begin{align}
    [\Omega(g)]_{(cd)(ab)} = \frac{\eta^{*}_g}{2} &( [\tilde{D}_{\gamma}(g)]_{ca} [\tilde{D}_{\gamma}(g)]_{db} \nonumber \\
    &- [\tilde{D}_{\gamma}(g)]_{da} [\tilde{D}_{\gamma}(g)]_{cb} ). \label{eq:rho_g2}
\end{align}
Thus, the character of $\rho^{(ab)}$ is given by the trace of Eq.~(\ref{eq:rho_g2}) as
\begin{align}
    \chi^{\Omega}_g &= \frac{\eta^{*}_g}{2}\left\{ \text{tr}[\tilde{D}_{\gamma}(g)]^2-\text{tr}[\tilde{D}_{\gamma}^2(g)]\right\}, \label{eq:chi0}
\end{align}
which determines IRs of $\rho^{(ab)}$ after performing the irreducible decomposition.
In general, short representations are a reducible representation, so it can be  decomposed into IRs of $G_0$ for type (I) groups and IRs of $H_0$ for type (III) groups. 
Then, $\chi^{\Omega}_g$ can be evaluated by using the character table or the explicit form of $\tilde{D}_{\gamma}(g)$. 

In type-(III) groups, an antiunitary operator $\tilde{D}(Th)$ imposes an additional constraint on $\rho^{(ab)}$. To illustrate this, we use the approach based on topological invariants~\cite{yamazaki094508}. Another approach based on group theory is discussed in Appendix~\ref{app:representation}. Suppose that MFs are associated with a 1D winding number $W[h]$ (labels of the subspace are omitted for simplicity). The coupling between MFs and $\tilde{O}$ is possible only if $\tilde{O}$ makes $W[h]$ ill-defined, which breaks the chiral symmetry, $[\tilde{O},\Gamma(Th)]=0$. Thus, when the trace part of Eq.~(\ref{eq:multipole}) is nonzero,
\begin{align}
    \Gamma(Th) \rho^{(ab)} \Gamma^{\dagger}(Th)  = \rho^{(ab)} \label{eq:chi_chiral}
\end{align}
must hold. Since $\tau_x \rho^{(ab) \ast} \tau_x = -\rho^{(ab)}$, Eq.~(\ref{eq:chi_chiral}) becomes
\begin{align}
     \tilde{D}(Th) \rho^{(ab) \ast} \tilde{D}^{\dagger}(Th)  = -\rho^{(ab)}. \label{eq:chi_Tg}
\end{align}
Similarly, $\tilde{O}$ satisfies $\tau_x \tilde{O}^{\ast} \tau_x = -\tilde{O}$, so we obtain
\begin{align}
    \tilde{D}(Th) \tilde{O}^{\ast} \tilde{D}^{\dagger}(Th) = -\tilde{O},
 \end{align}
which lead to, from Eq.~(\ref{eq:magnetic_coupling}) and (\ref{eq:strain_coupling}),
\begin{align}
    f(\bm{B}) = -f(-h\bm{B}) , \ \  f(\bar{u}) = -f(h\bar{u}h^{T}). \label{eq:chiral_f}
\end{align}

In short, the representation of $\rho^{(ab)}$ is determined from Eqs.~(\ref{eq:chi0}) and (\ref{eq:chi_Tg}), which only depend on magnetic point groups, IRs of the gap function, topological invariants that support MFs.

 \begin{table*}
	\caption{Electromagnetic responses of two MFs protected by type-(I) groups. The third column shows the IR of $\rho^{(12)}$, which is uniquely determined from $M_0^{\text{(I)}}$, the IR of $\Delta$, and topological invariants. The fourth column shows possible topological invariants, where $\nu[g]_{j,j'} =1$ is the shorthand notation for $\nu[g]_{j} =\nu[g]_{j'}=1$. When $\nu[g]_{j} =\nu[g']_{j} =1$, it means that an MF is protected by both $\nu[g]_{j}$ and $\nu[g']_{j}$. The fifth and sixth columns represent the symmetry-adopted forms of $f(\bm{B})$ and $f(\bar{u})$ in the leading order, determined solely from Eqs.~(\ref{eq:chi0}).  The Hermann-Mauguin notation for magnetic point group is adopted, and the labels of IRs are in the Mulliken notation~\cite{Bradley72}. }
		\begin{tabular}{lccccc}
             \hline \hline
			$M_0^{\text{(I)}}$ & IR of $\Delta$ & IR of $\rho^{(12)}$ & Topological invariants & $f(\bm{B})$ & $f(\bar{u})$
			\\
			\hline
			$2$ & $B$ & $B$ & $\nu[2_z]_{1,2}=1$ & $B_x, B_y$ & $u_{zx}$, $u_{yz}$ 
			\\
                $4$ & $^1 E$ & $B$ & $\nu[4_z]_{1,3}=1$ & $B_x^2-B_y^2, B_xB_y $ & $u_{xx}-u_{yy}$, $u_{xy}$ 
			\\
                $4$ & $^2 E$ & $B$ & $\nu[4_z]_{2,4}=1$ & $B_x^2-B_y^2, B_xB_y $ & $u_{xx}-u_{yy}$, $u_{xy}$ 
			\\
                $6$ & $B$ & $B$ & $\nu[6_z]_{2,5}=1$ & $B^3_x-3B_xB^2_y, B^3_y-3B_yB^2_x$ &  
			\\
                $6$ & $^1 E_2$ & $B$ & $\nu[6_z]_{1,4}=1$ & $B^3_x-3B_xB^2_y, B^3_y-3B_yB^2_x$ &   
			\\
                $6$ & $^2 E_2$ & $B$ & $\nu[6_z]_{3,6}=1$ & $B^3_x-3B_xB^2_y, B^3_y-3B_yB^2_x$ &   
			\\
   			$m$ & $B$ & $B$ & $\nu[m_{zx}]_{1,2}=1$ & $B_x, B_z$ & $u_{xy},u_{yz}$
			\\
         	$mm2$ & $A_2$ & $A_2$ & $\nu[m_{zx}]_{1,2}=\nu[m_{yz}]_{1,2}=1$ & $B_z$ & $u_{xy}$
			\\
                $mm2$ & $B_1$ & $B_1$ & $\nu[2_z]_{1,2}=\nu[m_{yz}]_{1,2}=1$ & $B_y$ & $u_{zx}$
			\\
                $mm2$ & $B_2$ & $B_2$  & $\nu[2_z]_{1,2}=\nu[m_{zx}]_{1,2}=1 $ & $B_x$ & $u_{yz}$
			\\
                $3m$ & $A_2$ & $A_2$ & $\nu[m_{zx}]_{1,2}=1$ & $B_z$ & 
                \\
                $4mm$ & $A_2$ & $A_2$ & $\nu[m_{zx}]_{1,2}=\nu[m_{d}]_{1,2}=1$  & $B_z$ & 
			\\
                $6mm$ & $A_2$ & $A_2$ & $\nu[m_{zx}]_{1,2}=\nu[m_{yx}]_{1,2}=1$ & $ B_z$ &  
   			\\
                $6mm$ & $B_1$ & $B_1$ & $\nu[6_z]_{1,2}=\nu[m_{zx}]_{1,2}=1$ & $B_x^3-3B_x B_y^2$ &  
   			\\
                $6mm$ & $B_2$ & $B_2$ & $\nu[6_z]_{1,2}=\nu[m_{yz}]_{1,2}=1$ &  $B_y^3-3B_yB_x^2$ & 
   			\\  			
              \hline \hline
 \end{tabular}	
\label{classification_type1}
\end{table*}

 \begin{table*}
\caption{
 Electromagnetic responses of two MFs protected by type-(III) groups, where the gap function and $\rho^{(12)}$ are classified by the IRs of $H_0$. 
 The fifth column shows possible topological invariants, where $W[h]^g_{j,j'} =1$ is the shorthand notation for $|W[h]^g_{j}| =|W[h]^g_{j'}|=1$. When $|W[h]^g_{j}| =|W[h']^g_{j}| =1$, it means that an MF is protected by both $W[h]^g_{j}$ and $W[h']^g_{j}$. We extract the integer part from $W[4_z]^{2_z}_{1,2}$ and $W[6_z]_{1,3}^{3_z}$. The sixth and seventh columns represent the symmetry-adopted forms of $f(\bm{B})$ and $f(\bar{u})$ in the leading order, determined from Eqs.~(\ref{eq:chi0}) and (\ref{eq:chi_Tg}). The notations are the same as those in Table~\ref{classification_type1}.
 }
		\begin{tabular}{lcccccc}
             \hline \hline
			$M_0^{\text{(III)}}$ & $H_0$ & IR of $\Delta$ & IR of $\rho^{(12)}$ & Topological invariants & $f(\bm{B})$ & $f(\bar{u})$
			\\
	        \hline \hline
			$2'$ &$1$ & $A$ & $A$ & $W[2_z]=2$ & $B_z$ & $u_{zx},u_{yz}$
			\\
                $4'$ & $2$ & $A$ & $A$ & $W[4_z]^{2_z}_{1,2}=1$ & $B_z$ & $u_{xx}-u_{yy},u_{xy}$
			\\
                $4'$ & $2$ & $B$ & $B$ & $\nu[2_z]_{1,2}=1$ & $B_x, B_y$ & $u_{zx},u_{yz}$
			\\
                $6'$ & $3$ & $A_1$ & $A_1$ & $W[6_z]^{3_z}_2=2$ & $B_z$ & 
			\\
                $6'$ & $3$ & $A_1$ & $A_1$ & $W[6_z]^{3_z}_{1,3}=1$ & $B_z$ & 
			\\
      		$m'$ & $1$ & $A$ & $A$ & $W[m_{zx}]=2$ & $B_y$ & $u_{xy},u_{yz}$
			\\
               $m'm2'$ & $m$ & $A''$ & $A''$ & $W[2_z]^{m_{zx}}_{1,2}=1$ & $B_z$ & $u_{yz}$
			\\
               $m'm2'$ & $m$ & $A''$ & $A''$ & $W[m_{yz}]^{m_{zx}}_{1,2}=1$  & $B_x $ & $u_{xy}$
			\\
               $m'm2'$ & $m$ & $A''$ & $A'$ & $W[2_z]^{m_{zx}}_{1,2}=W[m_{yz}]^{m_{zx}}_{1,2}=1$ & $B_xB_z $ & $u_{zx}$
			\\
               $m'm'2$ & $2$ & $B$ & $B$ & $W[m_{zx}]^{2_z}_{1,2}=1$ & $B_y$ & $u_{yz}$
			\\
               $m'm'2$ & $2$ & $B$ & $B$ & $W[m_{yz}]^{2_z}_{1,2}=1$ & $B_x$ & $u_{zx}$
			\\
               $m'm'2$ & $2$ & $B$ & $A$ & $W[m_{zx}]^{2_z}_{1,2}=W[m_{yz}]^{2_z}_{1,2}=1$ & $B_xB_y $ & $u_{xy}$
			\\
               $3m'$ & $3$ & $A_1$ & $A_1$ & $W[m_{yz}]^{3_z}_2=2$ & $B^3_x-3B_xB^2_y$ &  
			\\
               $3m'$ & $3$ & $^1 E$ & $A_1$ & $W[m_{yz}]^{3_z}_1=2$ & $B^3_x-3B_xB^2_y$ &  
			\\
               $3m'$ & $3$ & $^2 E$ & $A_1$ & $W[m_{yz}]^{3_z}_3=2$ & $B^3_x-3B_xB^2_y$ &  
			\\
               $4'm'm$ &$mm2$& $A_2$ & $A_2$ & $W[4_z]^{2_z}_{1,2}=1$ & $B_z$ & $u_{xx}-u_{yy}$
			\\ 
               $4m'm'$ &$4$& $^1 E$ & $B$ & $W[m_{yz}]^{4_z}_{1,3}=1$ & $B_x B_y $ & $u_{xy}$
			\\  
               $4m'm'$ &$4$& $^1 E$ & $B$ & $W[m_{d}]^{4_z}_{1,3}=1$ & $B_x^2 -B_y^2 $ & $u_{xx}-u_{yy}$
			\\
               $4m'm'$ &$4$& $^1 E$ & $A$ & $W[m_{yz}]^{4_z}_{1,3}=W[m_{d}]^{4_z}_{1,3}=1$ & $B_xB_y(B_x^2 -B_y^2) $ & 
			\\
               $4m'm'$ &$4$& $^2 E$ & $B$ & $W[m_{yz}]^{4_z}_{2,4}=1$ & $B_x B_y $ & $u_{xy}$
			\\  
               $4m'm'$ &$4$& $^2 E$ & $B$ & $W[m_{d}]^{4_z}_{2,4}=1$ & $B_x^2 -B_y^2 $ & $u_{xx}-u_{yy}$
			\\
               $4m'm'$ &$4$& $^2 E$ & $A$ & $W[m_{yz}]^{4_z}_{2,4}=W[m_{d}]^{4_z}_{2,4}=1$ & $B_xB_y(B_x^2 -B_y^2) $ & 
			\\
                $6'mm'$ & $3m$ & $A_2$ & $A_2$ & $W[6_z]_{1,3}^{3_z}=1$& $B_z$ & 
                \\
                $6'mm'$ & $3m$  & $A_2$ & $A_2$ & $W[6_z]^{3_z}_{2}=2$ & $B_z$ & 
   			\\
                $6'mm'$ & $3m$  & $A_2$ & $A_2$ & $W[m_{zx}]^{3_z}_{2}=2$ & $B_y^3-3 B_y B_x^2$ &  
   			\\
                $6'mm'$ & $3m$  & $A_2$ & $A_1$& $W[6_z]^{3_z}_{2}=W[m_{zx}]^{3_z}_{2}=2$ & $B_z(B_y^3-3 B_y B_x^2)$ & 
   			\\
                $6m'm'$ & $6$ & $B$ & $B$ & $W[m_{yz}]_{2,5}^{6_z}=1$ & $B_x^3-3 B_x B^2_y$ & 
   			\\
                $6m'm'$ & $6$ & $B$ & $B$ & $W[m_{zx}]_{2,5}^{6_z}=1$ & $B_y^3-3 B_y B^2_x$ & 
   			\\
                $6m'm'$ & $6$ & $B$ & $A$ & $W[m_{yz}]_{2,5}^{6_z}=W[m_{zx}]_{2,5}^{6_z}=1$ & $(B_x^3-3 B_x B^2_y)(B_y^3-3 B_y B^2_x)$ & 
   			\\
                $6m'm'$ & $6$ & $^1 E_2$ & $B$ & $W[m_{yz}]_{1,4}^{6_z}=1$ & $B_x^3-3 B_x B^2_y$ &  
   			\\
                $6m'm'$ & $6$ & $^1 E_2$ & $B$ & $W[m_{zx}]_{1,4}^{6_z}=1$ & $B_y^3-3 B_y B^2_x$ & 
   			\\
                $6m'm'$ & $6$ & $^1 E_2$ & $A$ & $W[m_{yz}]_{1,4}^{6_z}=W[m_{zx}]_{1,4}^{6_z}=1$ & $(B_x^3-3 B_x B^2_y)(B_y^3-3 B_y B^2_x)$ & 
   			\\
                $6m'm'$ & $6$ & $^2 E_2$ & $B$ & $W[m_{yz}]_{3,6}^{6_z}=1$ & $B_x^3-3 B_x B^2_y$ &  
   			\\
                $6m'm'$ & $6$ & $^2 E_2$ & $B$ & $W[m_{zx}]_{3,6}^{6_z}=1$ & $B_y^3-3 B_y B^2_x$ &  
   			\\
                $6m'm'$ & $6$ & $^2 E_2$ & $A$ & $W[m_{yz}]_{3,6}^{6_z}=W[m_{zx}]_{3,6}^{6_z}=1$ & $(B_x^3-3 B_x B^2_y)(B_y^3-3 B_y B^2_x)$ &  
   			\\
              \hline \hline
 \end{tabular}	
\label{classification_type3}
\end{table*}

\subsection{Electromagnetic responses of two MFs}
We now apply Eqs.~(\ref{eq:chi0}) and (\ref{eq:chi_Tg}) to MFs protected by a magnetic point group. We consider two MFs, $|u_0^{(1)}\rangle$ and $|u_0^{(2)}\rangle$, at a high symmetry point in the surface BZ as a minimal set of MFs. As discussed in Sec.~\ref{sec:degeneracy}, there are two types of double degeneracy: (a) accidental degeneracy and (b) symmetry-enforced degeneracy. In (b), double degeneracy is enforced by the magnetic point group, so we only have a pair of MFs, whose short representation is an IR of the magnetic point group and is given by a rotation matrix of a spin-$J/2$ fermion. Since the antisymmetric product of a rotation matrix is equivalent to a spin-singlet representation, the character of $\chi^{\Omega}_g$ is simply given by
\begin{align}
    \chi^{\Omega}_g = \eta_g.
\end{align}
This relation is verified for both type-(I) groups ($g \in G_0$) and type-(III) groups ($g \in H_0$). 

On the other hand, the character of $\chi^{\Omega}_g$ in (a) depends on pairs of MFs, whose short representation is a reducible representation. The short representations depend on the magnetic point group and topological invariants. Comprehensive lists of all short representations are provided in Tables~\ref{tab:short1} and~\ref{tab:short3} in the Appendix. 

Magnetic responses of two MFs under type-(I) groups, which are determined solely from IRs of $\rho^{(12)}$, are summarized in Tables~\ref{classification_type1}, where the symmetry-adopted forms of $f(\bm{B})$ and $f(\bar{u})$ in the leading order are shown. Here, the first, second, and third order terms of 
$f(\bm{B})$ are called the magnetic dipole, quadrupole, and octapole responses, respectively. Breaking of TRS allows to have $\bm{B}$-even terms in $f(\bm{B})$, e.g., when $M_0^{\text{(I)}} = 4$, and IR of the gap function is $^1 E$, $f(\bm{B})$ in the leading order becomes
\begin{align}
    f^4_{^1 E}(\bm{B}) \sim \rho_{x^2-y^2} (B_x^2-B_y^2) + \rho_{xy} B_x B_y,
\end{align}
since IR of $\rho^{(12)}$ is $B$, and thus, the crystalline symmetry imposes $f(\bm{B})=-f(4_z \bm{B})$, where $\rho_{x^2-y^2}$ and $\rho_{xy}$ are a real coefficient.

In systems with type-(III) groups, antiunitary operators impose an additional constraint on $f(\bm{B})$ and $f(\bar{u})$ through Eq.~(\ref{eq:chiral_f}), when a 1D winding number is nonzero. 
Thus, an anisotropic magnetic response occurs even when no unitary symmetry operator exists. For example, when $M^{\text{(III)}}_0=2'$, $W[2_z]=2$ leads to the magnetic dipole response; namely, $f(\bm{B})$ up to the leading order is given by 
\begin{align}
    f^{2'}_A (\bm{B}) \sim B_z \label{eq:mr_2'}
\end{align}
since $f(\bm{B})=-f(-2_z\bm{B})$ from Eq.~(\ref{eq:chiral_f}). 

Moreover, some magnetic point groups allow multiple 1D winding numbers. In this case, the magnetic response depends on the values of multiple 1D winding numbers.
For instance, we consider $M_0^{\text{(III)}}=m'm'2$ and the $B$ IR of the gap function. In this case, we have $W[m_{zx}]^{2_z}_{1,2}$ and $W[m_{yz}]^{2_z}_{1,2}$. The 1D winding numbers $W[m_{zx}]^{2_z}_{1,2}=1$ and  $W[m_{yz}]^{2_z}_{1,2}=1$, respectively, lead to $f^{m'm'2}_{B,1}(\bm{B}) \sim B_y$ and $f^{m'm'2}_{B,2}(\bm{B}) \sim B_x$ since $f(\bm{B})$ satisfies 
\begin{align}
    &f(\bm{B})=-f(-m_{zx}\bm{B}), \label{eq:mm2_f1}\\  
    &f(\bm{B})=-f(-m_{yz}\bm{B}). \label{eq:mm2_f2}
\end{align}
On the other hand, when $W[m_{zx}]^{2_z}_{1,2}=W[m_{yz}]^{2_z}_{1,2}=1$, $f(\bm{B})$ satisfy Eq.~(\ref{eq:mm2_f1}) and (\ref{eq:mm2_f2}) simultaneously, whereby leading to the magnetic quadratic response,
\begin{align}
    f^{m'm'2}_{B,3}(\bm{B}) \sim B_x B_y.
\end{align}
In general, when more than two 1D winding numbers are nonzero, $f(\bm{B})$ tend to have higher order terms in the leading order.
The electromagnetic responses of two MFs in type (III) groups are summarized in Tables~\ref{classification_type3}, where the magnetic response depends on not only IR of $\rho^{(12)}$ but also the 1D winding numbers. 

We note that in the case of magnetic dipole responses, $\bm{B}$-even terms also exist as second order terms. Including the second order terms in Eq.~(\ref{eq:mr_2'}) results in 
\begin{align}
    f^{2'}_A (\bm{B}) \sim \rho_z B_z + \rho_{xz} B_x B_z + \rho_{yz} B_y B_z,
\end{align}
where the amplitude of $\rho_{xz}, \rho_{yz}$ depends on that of TRS breaking terms such as a magnetization term. When the TRS breaking term is larger than the energy scale of the pair potential, the second order terms become dominant, leading to a quadrupole-like magnetic response.

\section{Application to TRSB TSCs}
\label{sec:app}
We demonstrate the magnetic response of two MFs in two models of TRSB TSCs. One model is of UTe$_2$ with a chiral superconducting state. The other model is of UCoGe with coexisting superconductivity and ferromagnetic order. In both cases, point nodes appear in the superconducting gap and are accompanied by chiral Majorana edge modes in a surface BZ. We study the magnetic response of two MFs arising from the double degeneracy of the chiral Majorana edge modes at a high-symmetry point.

\subsection{UTe$_2$}
\label{subsec: UTe2}

We consider the heavy fermion superconductor UTe$_2$~\cite{Ran684,aoki2022}, a potential candidate for TRSB odd-parity superconductors. The superconducting state of UTe$_2$ exhibits several unconventional properties such as high upper critical field beyond the Pauli limit~\cite{aoki2019unconventional,nakamine2019,knebel2019,ran2019extreme}, re-entrant superconductivity~\cite{knebel2019,ran2019extreme}, and multiple superconducting phases under pressure and magnetic field~\cite{braithwaite2019multiple,ran2020enhancement,aoki2020,Hayes797}. Some experimental studies have reported properties of TRSB superconductors, including the existence of point nodes~\cite{metz2019,kittaka2020,Ishihara2966} and TRS breaking~\cite{Jiao523,Bae2644,Ishihara2966}. 
However, specific heat measurements on UTe$_2$ single crystal show only a single superconducting transition at ambient pressure~\cite{Cairns415602, Thomas224501, Rosa33, Girod121101}, and recent Kerr effect measurements show no evidence for a spontaneous TRS breaking~\cite{Ajeesh041019}. Thus, the possibility of TRSB superconducting states remains open to debate.

In the following, we study the magnetic response of MFs in a chiral superconducting state of UTe$_2$. Using the recently proposed tight-binding model of UTe$_2$~\cite{Shishidou104504}, we demonstrate the quadratic magnetic response of two MFs, which indicates a signature of spontaneous TRS breaking. 

\subsubsection{Classification of gap functions}
 We assume that the normal state is nonmagnetic and that the breaking of TRS originates from the gap function. 
The crystal symmetry of UTe$_2$ is $mmm$, so the crystal symmetry operators are given by
\begin{align}
    mmm=\{e,2_x,2_y,2_z,I,m_{xy},m_{zx},m_{yz}\},
\end{align}
where $I$ represents spatial inversion. Cooper pairs are formed by Bloch functions that respect the $mmm$ crystal symmetry. Possible odd-parity gap functions are classified by IRs of $mmm$, labeled as $A_u$, $B_{1u}$, $B_{2u}$, and $B_{3u}$. These are all 1D IRs, indicating that possible odd-parity chiral states are constructed from a mixture of different IRs, e.g., $A_u + i B_{1u}$. Such a mixed state may occur when two superconducting states with different IRs are accidentally degenerate. 

We now relate odd-parity chiral states to the IRs of type-(III) groups with $H \subset mmm$. All chiral states can be assigned to the IRs of magnetic point groups. For example, consider the $A_u + i B_{1u}$ state, where the gap function of the $A_u$ ($B_{1u}$) state is even (odd) under time-reversal operation, and the transformation under the $mmm$ symmetry operation follows its IR as described in Eq.~(\ref{eq:gapfunctiong}). Although the $A_u + i B_{1u}$ state breaks the time-reversal and $mmm$ crystal symmetries, it remains invariant under
\begin{align}
    m'm'm = \{e,2_z, I, m_{xy}; T2_x,T2_y,Tm_{zx}, Tm_{yz}\}, \label{eq:m'm'm}
\end{align}
up to a sign: the $A_u + i B_{1u}$ state is even (odd) under $2_z$ ($I$ and $m_{xy}$), so it belongs to the $A_u$ IR of $H=112/m$. In a similar manner, IRs of type-(III) groups are assigned for other chiral pairings. Note that the IRs of $H$ do not depend on the choice of the gauge of the gap function; for example, the $B_{1u} + iA_u$ state is the $A_u + i B_{1u}$ state with a different gauge choice, so it also belongs to the $A_u$ representation of $H=112/m$. Table~\ref{tab:chiral} summarizes the relation among chiral states, representations of magnetic point groups, and corresponding 2D magnetic point groups that are invariant on the $(xy)$, $(zx)$, and $(yz)$ surfaces. The magnetic point groups $m'mm'$ and $mm'm'$ are defined by
\begin{align}
    &m'mm' = \{e,2_y, I, m_{zx}; T2_z,T2_x,Tm_{xy}, Tm_{yz}\}, \\
    &mm'm' = \{e,2_x, I, m_{yz}; T2_y,T2_z,Tm_{zx}, Tm_{xy}\}, 
\end{align}
and the 2D magnetic point groups, e.g. $m2'm'$, follow the same classification as $m'm2'$ in Table~\ref{Tab:top_inv_type3}, with permutations of $x$, $y$, and $z$ such as $x \to y$, $y \to z$, and $z \to x$.  

 \begin{table}
	\caption{Classification of odd-parity chiral states, where $\Delta$ is classified by the IRs of $H$. The fifth, sixth, and seventh columns represent the 2D magnetic point groups preserved on the $(xy)$, $(zx)$, and $(yz)$ surfaces, respectively. The 2D magnetic point groups are represented as $m'm'2=\{e,2_z;Tm_{zx},Tm_{yz}\}$, $m2'm'=\{e,m_{yz};T2_y,Tm_{xy}\}$, $2'mm'=\{e,m_{zx};T2_x,Tm_{xy}\}$, etc.
  The underline of the 2D magnetic point groups indicates the presence of 1D winding numbers. There are two 1D winding numbers in all cases; for instance, in systems with the $A_u+iB_{1u}$ state and the $(zx)$ surface, we can define $W[2_y]_{1,2}^{m_{xy}}$ and $W[m_{yz}]_{1,2}^{m_{xy}}$. }
		\begin{tabular}{lcccccc}
            \hline \hline
		   & $M^{\text{(III)}}$ & $H$ & IR of $\Delta$ & $(yz)$ & $(zx)$ & $(xy)$ 
			\\
			\hline
            $A_u+i B_{1u}$ & $m'm'm$ & $112/m$ & $A_u$ & $\underline{2'm'm}$ & $\underline{m'2'm}$ & $m'm'2$
            \\
            $A_u+i B_{2u}$ & $m'mm'$ & $12/m1$ & $A_u$ & $\underline{2'mm'}$ & $m'2m'$ & $\underline{m'm2'}$ 
            \\
            $A_u+i B_{3u}$ & $mm'm'$ & $2/m11$ & $A_u$ & $2m'm'$ & $\underline{m2'm'}$ & $\underline{mm'2'}$
            \\
            $B_{2u}+i B_{3u}$ & $m'm'm$ & $112/m$ & $B_u$ & $2'm'm$ & $m'2'm$ & $\underline{m'm'2}$
            \\
            $B_{1u}+i B_{3u}$ & $m'mm'$ & $12/m1$ & $B_u$ & $2'mm'$ & $\underline{m'2m'}$ & $m'm2'$
            \\
            $B_{1u}+i B_{2u}$ & $mm'm'$ & $2/m11$ & $B_u$ & $\underline{2m'm'}$ & $m2'm'$ & $mm'2'$
			\\
            \hline \hline
 \end{tabular}	
\label{tab:chiral}
\end{table}

\subsubsection{Tight-binding model of UTe$_2$}
\begin{figure}[t]
\centering
    \includegraphics[scale=0.06]{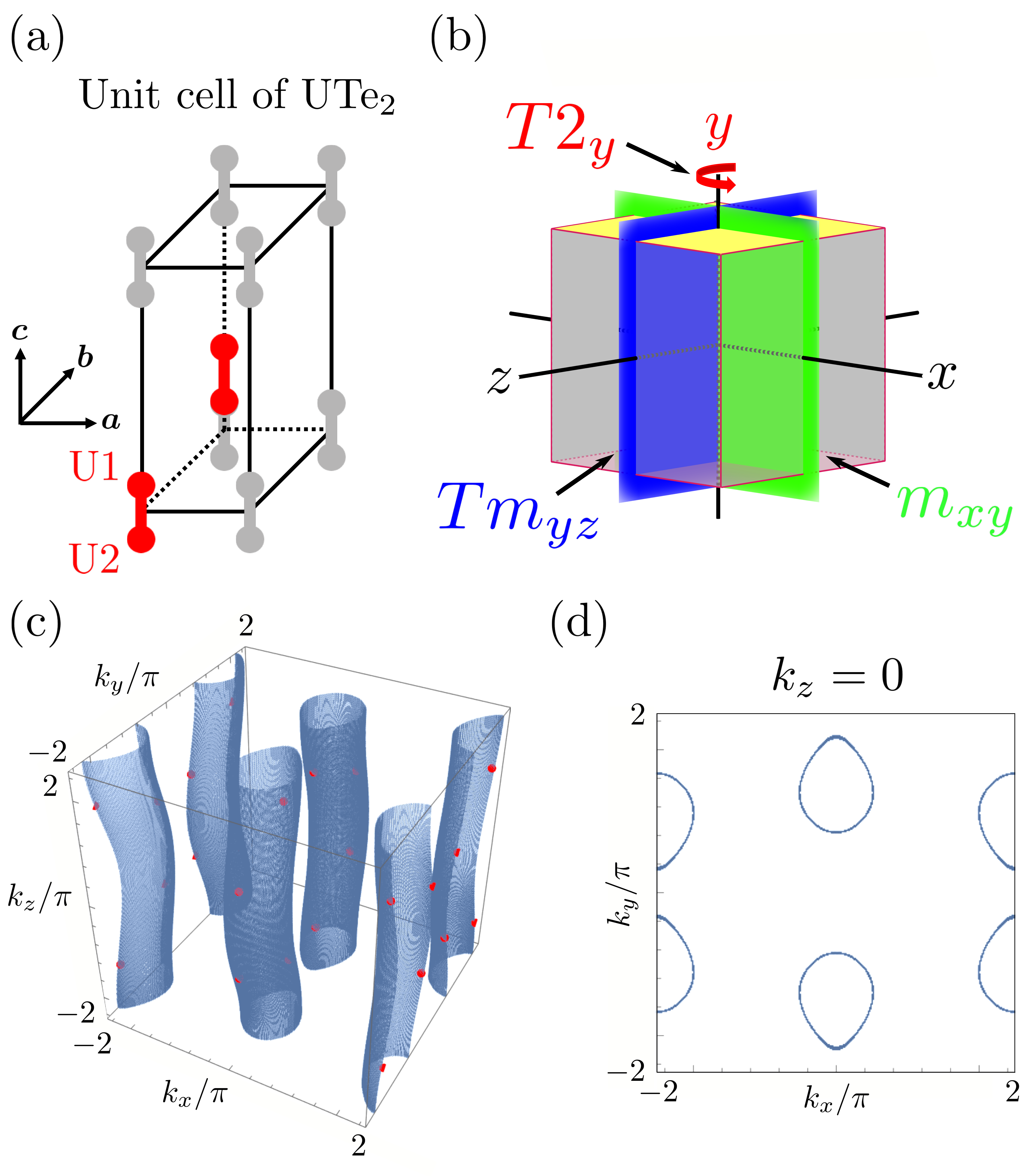}
    \caption{(a) Crystal structure of UTe$_2$, where a unit cell (red) includes two U atoms: U1 and U2. (b) Symmetry operations that preserve the $(zx)$ surface.  (c) Cylindrical Fermi surface of UTe$_2$. The position of superconducting point nodes are indicated by the red points. (d) The cross section of the Fermi surfaces at $k_z=0$. Each Fermi surface is doubly degenerate due to the inversion symmetry and TRS.}
    \label{Fig:UTe2-symmetry-operation}
\end{figure}

We consider a tight-binding model of UTe$_2$ proposed in Ref.~\onlinecite{Shishidou104504}, which takes into account the crystal symmetry of UTe$_2$ and sublattice structures of U atoms.
The tight-binding Hamiltonian is given by
\begin{align} 
\epsilon(\bm k) &= c(\bm k) + f_x(\bm k)\sigma_x + f_y(\bm k)\sigma_y + \bm{g(\bm{k})} \cdot \bm{s}\sigma_z, \label{normalpart1} 
\end{align}
with
\begin{subequations}
\label{normalpart2}
\begin{align} 
c(\bm k) &= 2t_1 \cos k_x + 2t_2 \cos k_y - \mu, \\ 
f_x(\bm k) &= t_3 + t_4\cos(k_x/2)\cos(k_y/2)\cos(k_z/2), \\
f_y(\bm k) &= t_5\cos(k_x/2)\cos(k_y/2)\sin(k_z/2), \\
g_x(\bm{k}) &= R_x \sin(k_y), \\
g_y(\bm{k}) &= R_y \sin(k_x), \\
g_z(\bm{k}) &= R_z \sin(k_x/2)\sin(k_y/2)\sin(k_z/2),
\end{align}
\end{subequations}
where $\bm{s}$ and $\bm{\sigma}$ denote the spin and sublattice degrees of freedom (U1 and U2) as shown in Fig.~\ref{Fig:UTe2-symmetry-operation} (a), respectively. The $t_{1-5}$ terms are the intra and inter sublattice hopping terms, $\mu$ is the chemical potential, and the $R_i$ terms $(i=x,y,z)$ are the spin-orbit coupling terms. 

For the superconducting state, we choose the $A_u + i B_{1u}$ state as an example.  
The explicit form of the gap function is given by
\begin{align}
    \Delta(\vv k) &= \Delta_{A_u}(\vv k) +  i \Delta_{B_{1u}}(\vv k), \nonumber \\
    &\equiv [\tilde{\Delta}_{A_u}(\vv k) +  i \tilde{\Delta}_{B_{1u}}(\vv k)] is_y \label{eq:gap_ute2}
\end{align}
with
\begin{align}
\tilde{\Delta}_{A_u}(\bm{k}) =&  \Delta^{A_u}_0 \sigma_y s_z + (\Delta^{A_u}_1 s_y+\Delta^{A_u}_2 \sigma_x s_y) j(\bm{k}) \nonumber \\
& + \Delta^{A_u}_3 s_y \sin(k_y), \\
\tilde{\Delta}_{B_{1u}}(\bm{k}) =& \Delta^{B_{1u}}_0 \sigma_z + (\Delta^{B_{1u}}_1 s_x+ \Delta^{B_{1u}}_2 \sigma_x s_x) j(\bm{k}) \nonumber \\ 
& + \Delta^{B_{1u}}_3 s_x \sin(k_y),
\end{align}
with $j(\bm{k}) \equiv \cos(k_x/2)\cos(k_z/2)\sin(k_y/2)$. Here, $\Delta_0$ is the on-site pair potential, $\Delta_1$ and $\Delta_2$ are the nearest neighbor pair potential, and $\Delta_3$ is the next nearest neighbor pair potential. All coefficients are real.

The BdG Hamiltonian (\ref{BdG22}) with the normal state (\ref{normalpart1}) and the gap function (\ref{eq:gap_ute2}) is invariant under the $m'm'm$ crystal symmetry operations (\ref{eq:m'm'm}). The matrix representations are given by
\begin{align}
&\tilde{D}(2_z)= i s_z \tau_z, \, \tilde{D}(m_{xy})= i\sigma_x s_z, \, \tilde{D}(I)= \sigma_x \tau_z,  \nonumber \\
&\tilde{D}(T2_x)= i\sigma_x s_z \tau_z, \, \tilde{D}(T2_y)= -\sigma_x, \nonumber \\ 
&\tilde{D}(Tm_{yz})= i s_z, \, \tilde{D}(Tm_{zx})= - \tau_z,
\end{align}
where the gauge of the gap function is chosen as $\eta_{T2_x}=1$. 

\subsubsection{Magnetic response of MFs}
We consider crystalline operators that remain surface invariant, which form 2D magnetic point groups for each surface (see Table~\ref{tab:chiral}). We can determine possible 1D topological invariants from the IRs of $\Delta$ in the 2D magnetic point groups compatible with those in $m'm'm$. For instance, on the $(zx)$ surface, the 2D magnetic point group is $m'2'm = \{e,m_{xy};T2_y, Tm_{yz} \}$ as shown in Fig.~\ref{Fig:UTe2-symmetry-operation} (b). Since the $A_u$ IR of $m'm'm$ is compatible with the $A''$ IR of $m'2'm$, we have two 1D winding numbers $W[2_y]_{1,2}^{m_{xy}}$ and $W[m_{yz}]_{1,2}^{m_{xy}}$ from Table~\ref{Tab:top_inv_type3}. Here, the 1D topological classification of $m'2'm$ is equivalent to that of $m'm2'$ under the exchange of the indices: $y \to z$ and  $z \to y$. 
As shown in Table~\ref{classification_type3}, the magnetic response of MFs depends on the 1D winding numbers. When either $W[2_y]_{1,2}^{m_{xy}}=1$ or $W[m_{yz}]_{1,2}^{m_{xy}}=1$, the magnetic dipole response appears, whereas when $W[2_y]_{1,2}^{m_{xy}}=W[m_{yz}]_{1,2}^{m_{xy}}=1$, the magnetic quadrupole response appears. 
 
\begin{figure}[t]
\vspace{5mm}
\centering
    \includegraphics[scale=0.38]{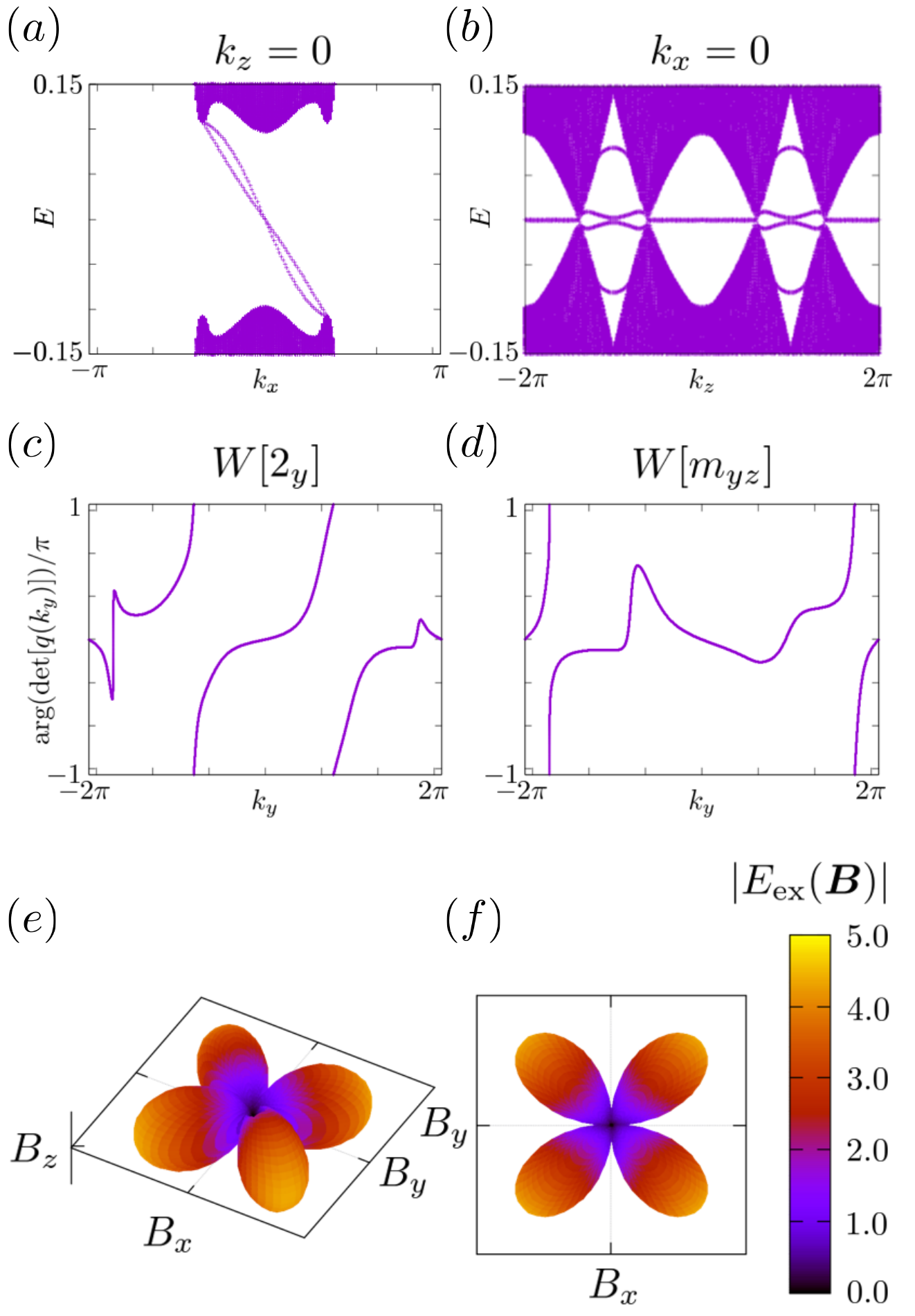}
    \caption{ In (a) and (b), we show the surface energy spectrum of the BdG Hamiltonian with the $A_u+iB_{1u}$ state for the $(zx)$ surface at (a) $k_z=0$ and (b) $k_x=0$. In (c) and (d), we show the change in the integrand in Eq.~(\ref{eq:re_1dwinding}) as a function of $k_y$ at $k_x=k_z=0$ for (c) $W[2_y]$ and (d) $W[m_{yz}]$. In (e) and (f), we illustrate the polar plot of the magnetic response of two MFs at $k_x=k_z=0$: (e) side view and (f) top view. The color represents the normalized magnitude of the energy gap $E_{\text{ex}}(\bm{B})$. }
    \label{Fig: energy-magnetic-response}
\end{figure}

To demonstrate the magnetic quadrupole response, we numerically diagonalize the BdG Hamiltonian with the open boundary condition in the $y$ direction. The BdG Hamiltonian with a finite size along the $y$ direction is represented as
\begin{align}
&H_{\text{open}}(\bm{k}_{\perp}) = \nonumber \\
&\sum_{m=0,1,2}\sum^{N_y-m}_{n=1} \bm{c}^{\dagger}_n(\bm{k}_{\perp}) H^{(m)}(\bm{k}_{\perp}) \bm{c}_{n+m}(\bm{k}_{\perp})
+ \text{h.c.} \label{eq:bdg_open}
\end{align}
where $\bm{k}_{\perp}=(k_x,k_z)$ and the system size $N_y$ along the $y$ direction is set to be $N_y = 800$. $H^{(0)}$, $H^{(1)}$, and $H^{(2)}$ correspond to the on-site terms, nearest neighbor hopping terms, and next-nearest neighbor hopping terms, respectively. We choose the parameters of the normal-state Hamiltonian to produce the cylindrical Fermi surface [see Fig.~\ref{Fig:UTe2-symmetry-operation} (c) and (d)]: $\mu=-1.8,\ t_1=-0.5,\ t_2=0.375,\ t_3=-0.7,\ t_4=0.65,\ t_5=-0.65,\ R_x=R_y=R_z=0.2$~\cite{Tei144517,Tei064516}, according to the recent measurement on the de Haas-van Alphen effect~\cite{Aoki083704, Eaton223}. The parameters of the gap function are chosen to be $\Delta^{A_u}_0=0.2$,  $\Delta^{A_u}_1=0.2$, $\Delta^{A_u}_2 =0.1$, $\Delta^{A_u}_3=0.12 $, $\Delta^{B_{1u}}_0=0.3$, $\Delta^{B_{1u}}_1=0.12$, $\Delta^{B_{1u}}_2=0.1$, and $\Delta^{B_{1u}}_3=0.04$. In these parameters, there are point nodes in the $k_x=0$ and $2 \pi$ planes. Figures~\ref{Fig: energy-magnetic-response} (a) and (b) show the energy spectrum of Eq.~(\ref{eq:bdg_open}), indicating a zero energy flat band state in a range of $k_z$ on the $k_x=0$ plane and two chiral Majorana edge modes on the $k_z=0$ plane. The chiral Majorana edge modes are attributed to nonzero Chern numbers of the point nodes. The zero energy flat band states on the $k_x=0$ plane occurs due to the chiral operator $\Gamma(Tm_{yz})$. Note that the model has other chiral Majorana edge modes in the $k_x=2\pi$ plane.

 The chiral Majorana edge modes are degenerate at $k_x=k_z=0$. We numerically calculate the 1D winding number at the high symmetry point using the modifined form:
\begin{align}
W[h] & = \frac{i}{4\pi} \int_{-2\pi}^{2\pi} dk_y \tr\left[ \Gamma(Th) H^{-1}(k_y) \pdv{H(k_y)}{k_y}\right] \nonumber \\
&=\int^{2\pi}_{-2\pi} \frac{dk_y}{2\pi} \frac{\partial}{\partial k_y}\text{arg}\{\text{det}[q(k_y)]\}, \label{eq:re_1dwinding}
\end{align}
where $h=2_y, m_{yz}$, $\Gamma(Th)^2=1$, and $H(k)$ is $4\pi$-periodic. The factor $q(k_y)$ is defined by
\begin{align}
U^{\dagger}_{\Gamma(Th)}H(0,k_y,0)U_{\Gamma(Th)} = \left[\begin{array}{cc}
0  & q(k_y) \\
q^{\dagger}(k_y)   & 0
\end{array}
\right],  \label{eq:def_q}
\end{align}  
where $U_{\Gamma(Th)} $ is a unitary matrix that diagonalizes the $\Gamma(Th)$. The 1D winding number is evaluated by the change in $\text{arg}(\text{det}[q(k_y)])$. We note that $W[h]=W^{m_{xy}}_{1}[h] +W^{m_{xy}}_{2}[h]$ for $h=2_y$ and $m_{yz}$ when we diagonalize $\Gamma(Th)$ and $H(k)$ simultaneously in terms of the eigenspaces of $m_{xy}$.
 Figures~\ref{Fig: energy-magnetic-response} (c) and (d) show the change in $\text{arg}(\text{det}[q(k_y)])$ as a function of $k_y$ for $W[2_y]$ and $W[m_{yz}]$, leading to $W[2_y]=W[m_{yz}]=2$. 

To verify the magnetic quadrupole response of the MFs at $k_x=k_z=0$, we add the Zeeman magnetic term,
\begin{align}
H_{\text{Z}}=\bm{c}^{\dagger}_1(\bm{k}_{\perp}) (\bm{B} \cdot \bm{s}) \bm{c}_{1}(\bm{k}_{\perp}), 
\label{surface-magnetic-fields}
\end{align} 
to Eq.~(\ref{eq:bdg_open}),
where we apply the Zeeman magnetic field only at the surface, i.e., $n=1$ and put $|\bm{B}|=0.1$~\footnote{We assume that the penetration depth is large enough so that the surface MFs feel the Zeeman magnetic field. In addition, we neglect the coupling between MFs and a screening field, which causes the Doppler shift in the energy dispersion of MFs~\cite{Chirolli2018} and does not change the magnitude of the energy gap of MFs.}. Figure~\ref{Fig: energy-magnetic-response} (e) and (f) show the energy gap of the MFs under the Zeeman magnetic field. We observe the magnetic quadrupole response, and the energy gap of the MFs is described as
\begin{align}
    E_{\text{ex}}(\bm{B}) \sim B_xB_y,
\end{align}
which is consistent with Table~\ref{classification_type3}. The similar results are obtained under strain in Appendix~\ref{sec:strain-response}.

Table~\ref{tab:chiral} presents the 1D topological classification for other chiral pairings, indicating that all odd-parity chiral pairing have multiple 1D winding numbers. Therefore, the magnetic quadrupole response also occurs for them in a certain surface.

\subsection{UCoGe}
\label{subsec: UCoGe}

This section discusses the ferromagnetic superconductor UCoGe~\cite{CANEPA1996225} as another example of the magnetic response of two MFs. UCoGe is a potential candidate for odd-parity superconductivity due to its ferromagnetism at ambient pressure~\cite{Huy067006, Aoki061011}. It offers a unique playground to study the interplay between ferromagnetism and superconductivity. The crystal symmetry of UCoGe is space group $Pnma$ (space group \# 53), with the little group at the $\Gamma$ point being $mmm$. In the ferromagnetic phase, the magnetization shows Ising-like anisotropy with the $c$-axis as a magnetic easy axis~\cite{Huy077002}, changing crystal symmetry changes from $Pnma$ to $Pn'm'a$, which coincides with $m'm'm$ at the $\Gamma$ point.  Additionally, NMR and Meissner measurements~\cite{Hattori066403} suggest that the ferromagnetic superconducting state has $A_u$ IR of $112/m$, where $112/m$ is the unitary subgroup of $m'm'm$.

We discuss the interplay between the magnetization and the magnetic response of MFs using a model of the ferromagnetic superconducting state of UCoGe with the $A_u$ IR. The breaking of time-reversal symmetry allows quadratic terms in $f(\bm{B})$ in addition to the linear term. When the magnitude of the magnetization is larger than that of the gap function, the magnetic quadratic response appears along with the magnetic dipole response, resulting in a highly anisotropic magnetic response.

\begin{figure}[t]
\centering
    \includegraphics[scale=0.06]{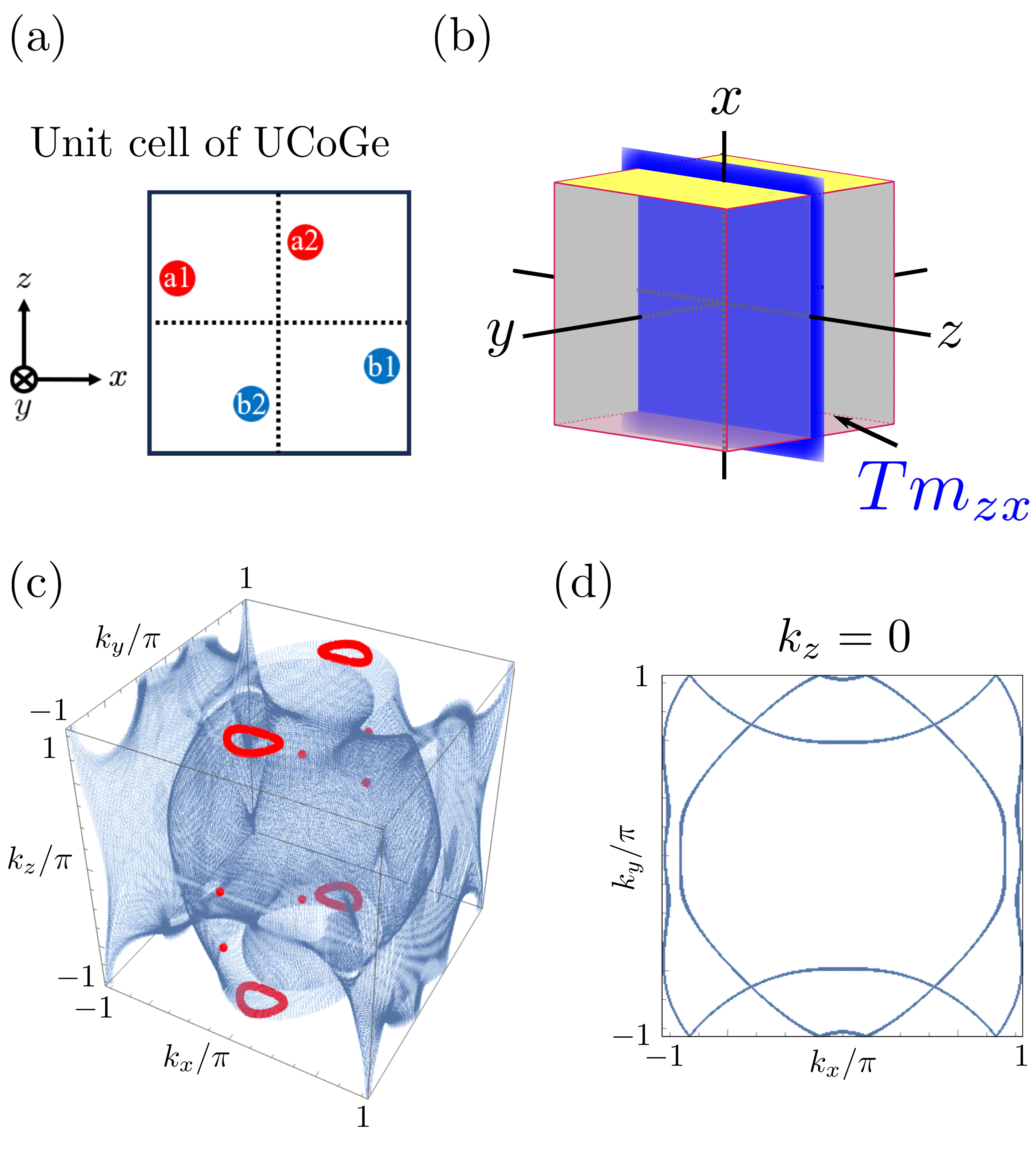}
    \caption{(a) Crystal structure of UCoGe, where a unit cell contains four U atoms ($a1$,$a2$,$b1$,$b2$). Their positions are given by $a1=(x-\frac{1}{2},-\frac{1}{4},z-\frac{1}{2}), \ a2=(x,-\frac{1}{4},1-z), \ b1=(\frac{1}{2}-x,\frac{1}{4},\frac{1}{2}-z),b2=(-x,\frac{1}{4},z-1)$, where $x=0.0101$ and $z=0.7075$~\cite{CANEPA1996225}. (b) Symmetry operations that preserve the $(yz)$ surface. (c) Cylindrical Fermi surface of UCoGe. The positions of superconducting point and line nodes are indicated by the red points and curves. (d) The cross section of the Fermi surface at $k_z=0$.}
    \label{Fig: UCoGe-symmetry-operation}
\end{figure}

\subsubsection{Tight-binding model of UCoGe} 
Since the unit cell has four U atoms positioned at $a1$, $a2$, $b1$, and $b2$ as shown in Fig. \ref{Fig: UCoGe-symmetry-operation} (a). 
The normal-state Hamiltonian is the $8 \times 8$ matrix including the spin, which is described as~\cite{Daido227001, Yoshida235105} 
\begin{align}
\epsilon(\bm{k}) =& c(\bm{k}) + t_1 \qty[\sigma_x + \lambda_1(k_x)] + \lambda_2(\bm{k})  \nonumber  \\ 
&+ \alpha\qty[\delta_{\alpha} \sin (k_x a_1) s_y - \sin (k_y a_2) s_x]\eta_z\sigma_z \nonumber \\ 
&+ \beta\qty[\sin(k_y a_2) s_z + \delta_{\beta}\sin(k_z a_3) s_y]\eta_z, \nonumber  \\
&+M_zs_z \label{eq:normal_ucoge} 
\end{align}
with
\begin{align}
c(\bm{k}) =&  2t'_1\cos(k_x a_1) + 2t_2 \cos(k_y a_2) \nonumber \\
&+ 2t_3 \cos(k_z a_3)  - \mu, 
\end{align}
where $s_i$, $\sigma_i$, and $\eta_i$ are the Pauli matrices in the space of the spin, sublattice $(1,2)$, and $(a,b)$, respectively. 
The third and forth lines in Eq.~(\ref{eq:normal_ucoge}) describe the spin-orbit coupling terms and mean-field magnetization term, respectively. $a_i$ ($i=1,2,3$) are the lattice constants. $\lambda_1(\bm{k})$ and $\lambda_2(\bm{k})$ are defined by 
\begin{align}
\lambda_1(\bm{k}) 
&= 
\left[ 
\begin{array}{cccc}
0 & e^{-ik_x a_1} & 0 & 0 \\
e^{ik_x a_1} & 0 & 0 & 0 \\
0 & 0 & 0 & e^{ik_x a_1} \\ 
0 & 0 & e^{-ik_x a_1} & 0
\end{array} 
\right]_{\eta \otimes \sigma},
\\
\lambda_2(\bm{k}) 
&=
\left[ 
\begin{array}{cccc}
0 & 0 & v_1(\bm{k}) & 0 \\
0 & 0 & 0 & v_2(\bm{k}) \\
v^{*}_1(\bm{k}) & 0 & 0 & 0 \\ 
0 & v^{*}_2(\bm{k}) & 0 & 0
\end{array} 
\right]_{\eta \otimes \sigma},
\end{align}
with $v_1(\bm{k}) = e^{-i k_x a_1}(1+e^{-ik_y a_2})(t_{ab}+e^{ik_z a_3}t'_{ab})$ and $v_2(\bm{k}) = e^{ik_z a_3}(1+e^{-ik_y a_2})(t_{ab}+e^{-ik_z a_3}t'_{ab})$~\footnote{The definition of the Fourier transformation in Eq.~(\ref{eq:normal_ucoge}) differs from that in Eq.~(\ref{BdG22}). In the basis of Eq.~(\ref{eq:normal_ucoge}), the crystalline symmetry operator depends on the momentum due to the sublattice degrees of freedom.}. The suffix implies the matrix representation in the $\eta \otimes \sigma$ space. 

For the gap function, we consider the $A_u$ state $\Delta(\bm{k}) = \tilde{\Delta}(\bm{k}) is_y$ with
\begin{align}
\tilde{\Delta} (\bm{k}) =& \Delta_0 \sin(k_x)s_x  \nonumber \\
                        & + \Delta_1 [\sin(k_y)s_y + \sin(k_z)s_x \sigma_z]. \label{eq:gapfn_ucoge}
\end{align}
In the absence of magnetization, the model retains the space group symmetry $Pnma$ and realizes topological crystalline superconductivity~\cite{Daido227001,Yoshida235105}. This state hosts two spinful MFs on a surface, which exhibit a magnetic quadrupole-like response~\cite{yamazaki094508}. It should be noted that a magnetic quadrupole-like response is allowed in time-reversal-invariant TSCs only when there are more than four MFs~\cite{yamazaki073701}.

 When the magnetization term is included, it realizes a point node superconductor similar to the chiral superconducting states of UTe$_2$. In the following, we consider the magnetic response of chiral Majorana edge modes with doubly degenerate MFs at a high-symmetry point.

\subsubsection{Magnetic response of MFs}
 There are point nodes in the intersection between the Fermi surfaces and $k_y=0$ plane as shown in Figure~\ref{Fig: UCoGe-symmetry-operation} (c), indicating that chiral Majorana edge modes emerge on the $(yz)$ surface.   Figure~\ref{Fig: UCoGe-symmetry-operation} (b) shows the symmetry operations that are preserved in the $(yz)$ surface. Only $Tm_{zx}$ remains invariant due to the nonsymmorphic nature~\footnote{The generators of the space group $Pnma$ are $g_1=\{2_x|\hat{\bm{x}}/2+\hat{\bm{y}}/2+\hat{\bm{z}}/2\}$,  $g_2=\{2_y|\hat{\bm{y}}/2\}$, $g_3=\{2_z|\hat{\bm{x}}/2+\hat{\bm{z}}/2\}$, $g_4=\{m_{yz}|\hat{\bm{x}}/2+\hat{\bm{y}}/2+\hat{\bm{z}}/2\}$,  $g_5=\{m_{zx}|\hat{\bm{y}}/2\}$, and $g_6=\{m_{xy}|\hat{\bm{x}}/2+\hat{\bm{z}}/2\}$, where $\{p|\bm{t}\}$ represents an element of space group: $p$ is a point group and $\bm{t}$ is a translation. On the $(yz)$ surface, we only have $g_5$ because the point group operations of $g_2$, $g_3$, and $g_4$ are not compatible with the surface, and $g_1$ and $g_6$ includes the translation in the $x$ direction. }. 
Since the 2D magnetic point group is $m'$, we have only the 1D topological invariant $W[m_{zx}]$.  The 1D winding number is defined by Eq.~(\ref{eq:re_1dwinding}) with
the $Tm_{zx}$ symmetry operator represented as
\begin{align}
\tilde{D}_{\bm{k}}(Tm_{zx}) = e^{-ik_z}
\left[ 
\begin{array}{cccc}
e^{-ik_y} & 0 & 0 & 0 \\
0 & e^{-ik_y} & 0 & 0 \\
0 & 0 & 1 & 0 \\ 
0 & 0 & 0 & 1
\end{array} 
\right]_{\eta \otimes \sigma} \tau_z,
\label{para}
\end{align}
where the momentum dependence comes from the sublattice degrees of freedom. 

\begin{figure}[t]
\vspace{5mm}
\centering
    \includegraphics[scale=0.38]{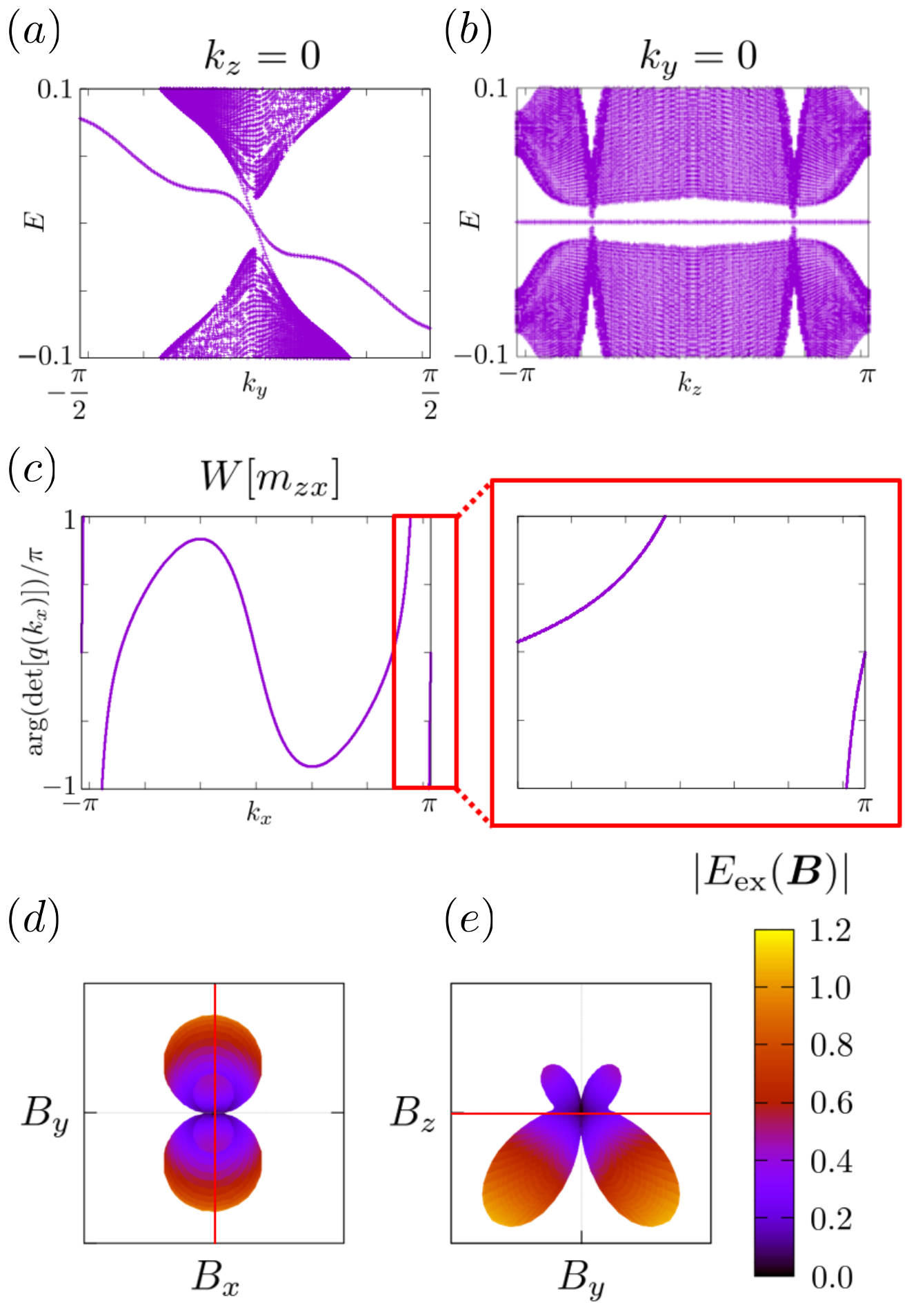}
    \caption{ In (a) and (b), we display the surface energy spectrum of the BdG Hamiltonian with the $A_u$ state for the $(yz)$ surface at (a) $k_z=0$ and (b) $k_y=0$. In (c), we depict the change in the integrand in Eq.~(\ref{eq:re_1dwinding}) with the $Tm_{zx}$ symmetry operator (\ref{para}) as a function of $k_x$ at $k_y=k_z=0$. The right panel shows the zoom around the $k_x=\pi$. In (d) and (e), we illustrate the polar plot of the magnetic response of two MFs at $k_y=k_z=0$: (d) $B_x-B_y$ plane and (e) $B_z-B_y$ plane. The color shows the normalized magnitude of the energy gap $E_{\text{ex}}(\bm{B})$.}
    \label{Fig: energy-magnetic-response-UCoGe}
\end{figure}

We numerically diagonalize the BdG Hamiltonian with Eq.~(\ref{eq:normal_ucoge}) and (\ref{eq:gapfn_ucoge}), where we set the lattice constants to be $1$ and choose  a cylindrical Fermi surface as shown in Figure~\ref{Fig: UCoGe-symmetry-operation} (c) and (d).  The parameters in the normal-state Hamiltonian are $(t_1, t_2, t_3, t_{ab}, t'_{ab}, \mu, t'_1, \alpha, \delta_{\alpha}, \beta, \delta_{\beta}) = (1, 0.2, 0.1, 0.5, 0.1, 0.55, 0.1, 0.3, 0.5, 0.3, 0.5)$~\cite{Fujimori174503, Fujimori062001,Daido227001}. The magnitudes of the magnetization term and superconducting order parameter are chosen to be $M_z=1.4$, $\Delta_0=0.4$, and $\Delta_1=0.2$. It is assumed that the magnetization does not significantly change the shape of the Fermi surface, as indicated by de Haas-van Alphen measurements~\cite{Leenen07024}.

The energy spectrum of the BdG Hamiltonian with the open boundary condition for the $x$ direction is shown in Figure~\ref{Fig: energy-magnetic-response-UCoGe} (a) and (b). There are two chiral Majorana edge modes with double degeneracy at $k_y=k_z=0$ [Fig.~\ref{Fig: energy-magnetic-response-UCoGe} (a)], forming the zero energy flat band states along the $k_z$ line [Fig.~\ref{Fig: energy-magnetic-response-UCoGe} (b)]. Figure~\ref{Fig: energy-magnetic-response-UCoGe} (c) shows the numerical calculation of the 1D winding number at $k_y=k_z=0$, leading to $W[m_{zx}] =2$. 

We demonstrate the magnetic response of MFs at $k_x=k_z=0$ by adding the magnetic Zeeman term as in Eq.~(\ref{surface-magnetic-fields}). As shown in Fig.~\ref{Fig: energy-magnetic-response-UCoGe} (d) and (e), we observe the highly anisotropic magnetic response that has the magnetic quadrupole response as a dominant contribution. The energy gap at the MFs is described as
\begin{align}
  E_{\text{ex}}(\bm{B}) \sim \rho_{y}B_y+\rho_{xy} B_xB_y+\rho_{yz} B_zB_y, 
\end{align}
where $\rho_y \gg \rho_{xy}$ and $\rho_{yz} > \rho_y$, indicating that the large $\rho_{yz}$ is induced by the magnetization term.

\section{Conclusion}
\label{sec:conclusion}
We established the multipole theory for MFs in TSCs with broken TRS and classified possible electromagnetic structures for two MFs protected by magnetic point groups, particularly type (I) and (III) groups as symmetry groups without TRS. By identifying possible pairs of MFs formed at a high-symmetry point in a surface BZ based on topological and group theoretical arguments, we found that the degeneracy of MFs falls into two categories: (a) accidental degeneracy and (b) symmetry-enforced degeneracy. We determined the electromagnetic structure of two MFs and associated symmetry-allowed coefficients: $f(\bm{B})$ and $f(\bar{u})$ based on the magnetic point groups, IR of the gap functions, and types of the degeneracy. As a result, we found that the electromagnetic structure of two MFs in type (b) shares the same IRs with the gap functions, indicating that pairing symmetries can be measured through the responses of two MFs to an external field. On the other hand, two MFs in type (a) do not have a direct relation to the gap functions, but exhibit time-reversal-even higher-order magnetic responses unique to TSCs with broken TRS, such as the magnetic quadrupole response. This opens a way to detect a spontaneous TRS breaking in TSCs.

We demonstrated the magnetic response of two MFs in the odd-parity chiral superconductor UTe$_2$ and the ferromagnetic superconductor UCoGe. For UTe$_2$, we classified possible odd-parity chiral pairing states under magnetic point groups and predicted that all pairing states have the magnetic quadrupole response of two MFs on a certain surface, as shown in Fig.~\ref{Fig: energy-magnetic-response} using the $A_u+iB_{1u}$ pairing state as an example. Furthermore, we studied the coexistence of ferromagnetic order and superconductivity in UCoGe as another example. We showed that the influence of the ferromagnetic order on the magnetic response of two MFs leads to the magnetic quadrupole responses as shown in Fig. \ref{Fig: energy-magnetic-response-UCoGe}. 
 
Finally, we address potential experimental methods for detecting the magnetic structures of MFs. Our findings indicate that MFs exhibit anisotropic spin structures, which can be examined using surface-sensitive techniques. These techniques include tunneling spectroscopy on a surface under a magnetic field or with a ferromagnet attached \cite{tanaka95, fogelstrom97, tanaka02, tanuma02, tanaka09, tamura17}, spin-resolved tunneling spectroscopy \cite{Jeon772, Cornils197002}, and ferromagnetic resonance of a ferromagnet/superconductor junction \cite{inoue17, kato19, Ominato121405}.

\begin{acknowledgments}

The authors thank S. Ono, K. Shiozaki, and A. Yamakage for useful discussions.
Y. Y. was supported by Grant-in-Aid for JSPS Fellows (Grant No. 22J14452).
S. K. was supported by JSPS KAKENHI (Grants No. JP22K03478 and No. JP24K00557). 
This work was supported by JST CREST (Grant No. JPMJCR19T2).

\end{acknowledgments}

\appendix

\section{Wigner's test}
\label{app:wigner}
In this appendix, we review the Wigner's test to identify the effective Altland-Zirnbauer symmetry classes under crystalline symmetry groups~\cite{Bradley72,Wigner59,Herring37,Inui90,Shiozaki2022,Sumita134513}, which is useful for identifying the 1D topological invariant. In the following, we consider the BdG Hamiltonian on a 1D subspace of the BZ, which is invariant under point group $G$ and antiunitary operator $T$, where $T$ need not be the time-reversal operation. The full symmetry group including the particle-hole operator $C$ is given by
\begin{align}
 \mathcal{G} = G + TG + CG + \Gamma G,
\end{align}
where $\Gamma \equiv TC$.
On the 1D subspace, the BdG Hamiltonian commutes with $\tilde{D}(g)$ $(g \in G)$, so they can be decomposed into 
\begin{align}
   \tilde{D}(g) = \oplus_{\alpha} D^{\alpha}(g), \ \  H(k) = \oplus_{\alpha} H^{\alpha}(k),
\end{align}
where $\alpha$ is the label of IRs of $G$, and $D^{\alpha}(g)$ is the $\alpha$ IR of $\tilde{D}(g)$.
The Wigner's test specifies an Altland-Zirnbauer symmetry class of $H^{\alpha}(k)$, and determines a possible 1D topological invariant of $H^{\alpha}(k)$.
For the Wigner's test of $H^{\alpha}(k)$, we calculate three indices $(W^T, W^C, W^{\Gamma})$ defined as follows:
\begin{align}
 &W_{\alpha}^{T} \equiv \frac{1}{|G|} \sum_{g \in G} 
  \tr [D^{\alpha}((Tg)^2)]=\pm 1, 0,
 \label{eq:WT} \\
 &W_{\alpha}^{C} \equiv \frac{1}{|G|} \sum_{g \in G}  \tr [D^{\alpha}((Cg)^2)]=\pm 1, 0,
 \label{eq:WC} \\
& W_{\alpha}^{\Gamma} \equiv \frac{1}{|G|} \sum_{g \in G}  \tr[D^{\alpha}(\Gamma^{-1} g \Gamma)]^{\ast} \tr[D^{\alpha}(g)]
=1,0,
\label{eq:WG} 
\end{align}
where $|G|$ represents the number of elements in $G$.
The indices $(W_{\alpha}^{T},W_{\alpha}^{C},W_{\alpha}^{\Gamma})$ identify the Altland-Zirnbauer symmetry class of $H^{\alpha}(k)$, as shown in Table~\ref{tab:EAZ}.
From the Altland-Zirnbauer symmetry class, we can specify the possible 1D topological invariant. For type-(I) groups, there is no antiunitary operator, so the AZ symmetry class is identified solely by $W_{\alpha}^{C}$. The $\mathbb{Z}_2$ topological invariant is possible when $H^{\alpha}(k)$ belongs to class D. For type-(III) groups, we regard an anti-unitary operator as $T$. When we obtain classes AIII and BDI, we have the $\mathbb{Z}$ topological invariant. The 1D topological classifications for type-(I) and type-(III) groups are summarized in Tables~\ref{Tab:top_inv_type1} and \ref{Tab:top_inv_type3}.

\begin{table}[tb]
\caption{
Relationship between the indices $(W_{\alpha}^{T},W_{\alpha}^{C},W_{\alpha}^{\Gamma})$, Altland-Zirnbauer (AZ) symmetry classes, and 1D topological invariants.
}
\label{tab:EAZ}
\begin{tabular}{ccccc}
\hline\hline
$W_{\alpha}^{T}$ & $W_{\alpha}^{C}$& $W_{\alpha}^{\Gamma}$ &AZ class & 1D topo.\\
\hline 
$0$ & $0$ & $0$ & A & $0$\\
$0$ & $0$ & $1$ & AIII &  $\mathbb{Z}$\\
$1$ & $0$ & $0$ & AI &  $0$\\
$1$ & $1$ & $1$ & BDI &  $\mathbb{Z}$ \\
$0$ & $1$ & $0$ & D &  $\mathbb{Z}_2$\\
$-1$ & $1$ & $1$ & DIII & $\mathbb{Z}_2$\\
$-1$ & $0$ & $0$ & AII &  $0$\\
$-1$ & $-1$ & $1$ & CII & $2 \mathbb{Z}$\\
$0$ & $-1$ & $0$ & C &  $0$\\
$1$ & $-1$ & $1$ & CI &  $0$\\
\hline\hline
\end{tabular} 
\end{table}  

\section{Representation of $\rho^{(ab)}$ in type (III) groups}
\label{app:representation}

\begin{table}
	\caption{Short representations of two MFs for type-(I) groups. The labels of the magnetic point groups are in the Hermann-Mauguin notation. The short representations are represented by the representations of the magnetic point groups in the Bilbao Crystallographic Server~\cite{Elcoro}, where $\bar{\Gamma}_i$ denotes the $i$th double-valued IR of the little group at the $\Gamma$ point. Here, $2\bar\Gamma_i$ is the shorthand notation of $\bar\Gamma_i \oplus \bar\Gamma_i$. $C_{\gamma}$ is the particle-hole operator in the Majorana basis, where $s_0$ is the $2\times 2$ identity matrix, and $s_i$ ($i=x,y,z$) are the Pauli matrices in the space of two MFs. }
		\begin{tabular}{lcccc}
            \hline \hline
			$M_0^{\text{(I)}}$ & IR of $\Delta$ & Short rep. & $C_{\gamma}$ & Degeneracy
			\\
			\hline
		$2$ & $B$  & $\bar\Gamma_{3} \oplus \bar\Gamma_{4}$ & $s_0,s_z$ & (a)
   		\\
     	$2$ & $B$  & $2\bar\Gamma_{3}, 2\bar\Gamma_{4}$ & $s_0,s_z,s_x$ & (a)
   		\\
     	$4$ & $^1 E$  & $\bar{\Gamma}_5 \oplus \bar{\Gamma}_6$ & $s_0, s_z$ & (a)
   		\\
     	$4$ & $^2 E$  & $\bar{\Gamma}_7 \oplus \bar{\Gamma}_8$ & $s_0, s_z$& (a)
   		\\
            $6$ & $B$ & $\bar\Gamma_7 \oplus \bar\Gamma_8$ & $s_0,s_z$& (a)
            \\
            $6$ & $^1 E_2$ & $\bar\Gamma_9 \oplus \bar\Gamma_{10}$ & $s_0,s_z$& (a)
            \\
            $6$ & $^2 E_2$ & $\bar\Gamma_{11} \oplus \bar\Gamma_{12}$ & $s_0,s_z$& (a)
            \\
   		$m$ & $A''$  & $\bar\Gamma_{3} \oplus \bar\Gamma_{4}$  & $s_0,s_z$ & (a)
            \\
            $m$ & $A''$  & $2\bar\Gamma_{3}, 2\bar\Gamma_{4}$  & $s_0,s_z,s_x$ & (a)
            \\
            $mm2$ & $A_2$  & $\bar\Gamma_5$ & $s_0$ &(b)
			\\
            $mm2$ & $B_1$  & $\bar\Gamma_5$ & $s_z$ & (b)
			\\
            $mm2$ & $B_2$  & $\bar\Gamma_5$ & $s_x$ & (b)
            \\
            $3m$ & $A_2$ & $\bar\Gamma_4 \oplus \bar\Gamma_5$ & $s_0,s_z$ & (a) 
            \\
            $3m$ & $A_2$ & $2\bar\Gamma_4, 2\bar\Gamma_5$ & $s_0,s_z,s_x$ & (a) 
            \\
            $3m$ & $A_2$ & $\bar\Gamma_6$ & $s_x$ & (b)
            \\
            $4mm$ & $A_2$ & $\bar\Gamma_6, \bar \Gamma_7$ & $s_x$& (b)
			\\
            $6mm$ & $A_2$ & $\bar\Gamma_7, \bar\Gamma_8, \bar\Gamma_9$ & $s_x$ & (b) 
            \\
            $6mm$ & $B_1$ & $\bar\Gamma_7$ & $s_0$ & (b)
   		\\
            $6mm$ & $B_2$ & $\bar\Gamma_7$ & $s_z$ & (b)
            \\
            \hline \hline
 \end{tabular}	
\label{tab:short1}
\end{table}

\begin{table}
	\caption{Short representations of two MFs for type-(III) groups. The notations are the same as those in Table~\ref{tab:short1}.}
		\begin{tabular}{lccccc}
            \hline \hline
			$M_0^{\text{(III)}}$ & $H_0$ & IR of $\Delta$ & Short rep. & $C_{\gamma}$ & Degeneracy
			\\
			\hline
		$2'$ & $1$ & $A$  & $2\bar\Gamma_2$  & $s_0$ & (a)
            \\
            $4'$ & $2$ & $A$ & $\bar\Gamma_3\bar \Gamma_4$  & $s_x$ & (b)
			\\
            $4'$ & $2$ & $B$ & $\bar\Gamma_3\bar \Gamma_4$ & $s_0,s_z$ & (b)
			\\
            $6'$ & $3$ & $A_1$ & $2\bar\Gamma_4$ & $s_0$ & (a)
            \\
            $6'$ & $3$ & $A_1$ & $\bar\Gamma_5 \bar\Gamma_6$  & $s_x$ & (b)
			\\
      	$m'$ & $1$ & $A$ & $2\bar\Gamma_2$ & $s_0$ & (a)
			\\
            $m'm2'$ & $m$ & $A''$ & $\bar\Gamma_3 \oplus \bar \Gamma_4$ & $s_0$ & (a)
			\\
            $m'm2'$ & $m$ & $A''$ & $\bar\Gamma_3 \oplus \bar \Gamma_4$ & $s_z$ & (a)
			\\
            $m'm2'$ & $m$ & $A''$  & $2\bar\Gamma_3, 2\bar \Gamma_4$ & $s_0$ & (a)
			\\
            $m'm'2$ & $2$ & $B$  & $\bar\Gamma_3 \oplus \bar \Gamma_4$ & $s_0$ & (a)
			\\
            $m'm'2$ & $2$ & $B$ & $\bar\Gamma_3 \oplus \bar \Gamma_4$ & $s_z$ & (a)
			\\
            $m'm'2$ & $2$ & $B$ & $2\bar\Gamma_3, 2\bar \Gamma_4$  & $s_0$ & (a)
			\\
            $3m'$ & $3$ & $A_1$ & $2\bar\Gamma_4$  & $s_0$ & (a)
			\\
            $3m'$ & $3$ & $^1 E$ & $2\bar\Gamma_5$  & $s_0$ & (a)
			\\
            $3m'$ & $3$ & $^2 E$ & $2\bar\Gamma_6$  & $s_0$ & (a)
			\\
            $4'm'm$ & $mm2$ & $A_2$ & $\bar\Gamma_5$  & $s_0$ & (b)
			\\
            $4m'm'$ & $4$ & $^1 E$ & $\bar\Gamma_5 \oplus \bar\Gamma_6$  & $s_0$ & (a)
			\\
            $4m'm'$ & $4$ & $^1 E$ & $\bar\Gamma_5 \oplus \bar\Gamma_6$  & $s_z$ & (a)
			\\
            $4m'm'$ & $4$ & $^1 E$ & $2\bar\Gamma_5, 2\bar\Gamma_6$  & $s_0$ & (a)
			\\
            $4m'm'$ & $4$ & $^2 E$ & $\bar\Gamma_7 \oplus \bar\Gamma_8$  & $s_0$ & (a)
			\\
            $4m'm'$ & $4$ & $^2 E$ & $\bar\Gamma_7 \oplus \bar\Gamma_8$  & $s_z$ & (a)
			\\
            $4m'm'$ & $4$ & $^2 E$ & $2\bar\Gamma_7, 2\bar\Gamma_8 $  & $s_0$ & (a)
			\\
            $6'mm'$ & $3m$ & $A_2$ &  $\bar\Gamma_6$ & $s_x$ & (b)
            \\
            $6'mm'$ & $3m$ & $A_2$ & $\bar\Gamma_4 \oplus \bar\Gamma_5$ & $s_0$ & (a) 
            \\
            $6'mm'$ & $3m$ & $A_2$ & $\bar\Gamma_4 \oplus \bar\Gamma_5$ & $s_z$ & (a)
            \\
            $6'mm'$ & $3m$ & $A_2$ & $2\bar\Gamma_4, 2\bar\Gamma_5$ & $s_0$ & (a)
            \\
            $6m'm'$ & $6$ & $B$ & $\bar\Gamma_7 \oplus \bar\Gamma_8$ & $s_0$ & (a)
            \\
            $6m'm'$ & $6$ & $B$ & $\bar\Gamma_7 \oplus \bar\Gamma_8$ & $s_z$& (a)
            \\
            $6m'm'$ & $6$ & $B$ & $2\bar\Gamma_7, 2\bar\Gamma_8$ & $s_0$  &(a)
            \\
            $6m'm'$ & $6$ & $^1 E_2$ & $\bar\Gamma_9 \oplus \bar\Gamma_{10}$ & $s_0$ & (a)
            \\
            $6m'm'$ & $6$& $^1 E_2$ & $\bar\Gamma_9 \oplus \bar\Gamma_{10}$ & $s_z$& (a)
            \\
            $6m'm'$ & $6$& $^1 E_2$ & $2\bar\Gamma_9, 2\bar\Gamma_{10}$ & $s_0$  &(a)
            \\
            $6m'm'$ & $6$ & $^2 E_2$ & $\bar\Gamma_{11} \oplus \bar\Gamma_{12}$ & $s_0$ & (a)
            \\
            $6m'm'$ & $6$ & $^2 E_2$ & $\bar\Gamma_{11} \oplus \bar\Gamma_{12}$ & $s_z$& (a)
            \\
            $6m'm'$ & $6$ & $^2 E_2$ & $2\bar\Gamma_{11}, 2\bar\Gamma_{12}$ & $s_0$  &(a)
            \\
            \hline \hline
 \end{tabular}	
\label{tab:short3}
\end{table}

In this appendix, we discuss the representation of $\rho^{(ab)}$ in type-(III) groups. For a given antiunitary operator $\tilde{D}(Th)$, we decompose $\rho^{(ab)}$ into~\cite{kobayashi224504}:
\begin{align}
    &\tilde{O} = \tilde{O}_{h,+}+\tilde{O}_{h,-}, \\
    &\rho^{(ab)} = \rho^{(ab)}_{h,+} + \rho^{(ab)}_{h,-}, 
\end{align}
with
\begin{align}
    &\tilde{O}_{h,\pm} \equiv \frac{\tilde{O} \pm \tilde{D}(Th)\tilde{O}^{T} \tilde{D}^{\dagger}(Th)}{2} , \\
    &\rho^{(ab)}_{h,\pm} \equiv \frac{\rho^{(ab)} \pm \tilde{D}(Th)\rho^{(ab)  T} \tilde{D}^{\dagger}(Th)}{2}. 
\end{align}
These operators satisfy
\begin{align}
   &\tilde{D}(Th) \tilde{O}_{h,\pm}^T \tilde{D}^{\dagger}(Th) = \pm \tilde{O}_{h,\pm}, \\
   &\tilde{D}(Th) \rho^{(ab) T}_{h,\pm} \tilde{D}^{\dagger}(Th) = \pm \rho^{(ab)}_{h,\pm},
\end{align}
where we have used $\tilde{D}(Th)\tilde{D}^{\ast}(Th) = \pm 1$ \footnote{For $g=4_z$ and $6_z$, we choose the basis that diagonalizes $\tilde{D}(2_z)$ and $\tilde{D}(3_z)$, respectively}. Using $\text{tr}(O\rho^{(ab)})=\text{tr}(O^T\rho^{(ab)T})$, the trace part of Eq.~(\ref{eq:multipole}) is decomposed into
\begin{align}
    \text{tr}(\tilde{O}\rho^{(ab)}) = \text{tr}(\tilde{O}_{h,+}\rho^{(ab)}_{h,+})+\text{tr}(\tilde{O}_{h,-}\rho^{(ab)}_{h,-}).
\end{align}
 Since $\tilde{O}^T = \tilde{O}^{\ast}$, the $\tilde{O}_{h,\pm}$ components give the even-and odd-parity components under $\tilde{D}(Th)$. The transformation of $\rho^{(ab) T}_{h,\pm}$ under a unitary operator $g$ is given by
 \begin{align}
     \tilde{D}(g) \rho^{(ab) T}_{h,\pm} \tilde{D}^{\dagger}(g) \equiv \sum_{cd} \rho^{(ab)}_{h,\pm} [\Omega(g)_{h,\pm}]_{(cd)(ab)},
 \end{align}
with 
\begin{align}
    [\Omega&(g)_{h,\pm}]_{(cd)(ab)} = \notag \\
    & \frac{\eta^{*}_g}{4} \Big\{ \Big( [\tilde{D}_{\gamma}(g)]_{ca} [\tilde{D}_{\gamma}(g)]_{db} - [\tilde{D}_{\gamma}(g)]_{da} [\tilde{D}_{\gamma}(g)]_{cb} \Big) \notag \\
    &\quad \mp \Big( [\Gamma_{\gamma} (Th) \tilde{D}_{\gamma}(g)]_{ca} [C_{\gamma}^{-1} \Gamma^{\ast}_{\gamma}(Th) C_{\gamma} \tilde{D}_{\gamma}(g)]_{db}  \notag \\
    &\quad - [(C_{\gamma}^{-1} \Gamma^{\ast}_{\gamma}(Th) C_{\gamma} \tilde{D}_{\gamma}(g)]_{da}[\Gamma_{\gamma} (Th) \tilde{D}_{\gamma}(g)]_{cb} \Big) \Big\}. \label{eq:rep_omega_pm}
\end{align}
where we have used the relation $ \rho^{(ab) T} =  \tau_x \rho^{(ab)} \tau_x$ and 
introduced the particle-hole operator and chiral operator in the Majorana basis:
\begin{align}
    &[C_{\gamma}]_{ab} \equiv \langle u_0^{(a)} | C u_0^{(a)} \rangle = \langle u_0^{(a)} | \tau_x | u_0^{(b) \ast} \rangle , \\
    &\{\Gamma_{\gamma}(Th)\}_{ab} \equiv \langle u_0^{(a)} | \tilde{D}(Th) \tau_x |u_0^{(b)} \rangle. 
\end{align}
Here, $C_{\gamma}$ satisfies the same multiplication law as $C$ such that
\begin{align}
    \tilde{D}_{\gamma} (g) C_{\gamma} = \eta_g  C_{\gamma} \tilde{D}^{\ast}_{\gamma} (g). 
\end{align}
Equation~(\ref{eq:rep_omega_pm}) does not depend on the phase factor $e^{i \phi}$ in $\Gamma(Th)$. Taking the trace of Eq.~(\ref{eq:rep_omega_pm}), the character of $\rho^{(ab)}_{h,\pm}$ is given by
\begin{align}
   \chi&^{\Omega_{\pm,h}}(g) =\notag \\
   & \frac{\eta^{*}_g}{4} \Big\{ \text{tr}[\tilde{D}_{\gamma} (g)]^2-\text{tr}[\tilde{D}_{\gamma}^2(g)] \notag \\
   &\quad \mp \Big( \text{tr} \big[\Gamma_{\gamma}(Th)\tilde{D}_{\gamma} (g)\big] \text{tr} \big[C_{\gamma}^{-1} \Gamma_{\gamma}^{\ast}(Th) C_{\gamma} \tilde{D}_{\gamma} (g) \big]\notag \\
   &\quad -\text{tr}\big[ C_{\gamma}^{-1} \Gamma_{\gamma}^{\ast}(Th) C_{\gamma} \tilde{D}_{\gamma} (g)\Gamma_{\gamma}(Th)\tilde{D}_{\gamma} (g) \big]\Big\}. \label{eq:chig}
\end{align}
When $h=e$, $\tilde{D}(Te)$ is the time-reversal operation. Using $C_{\gamma}^{-1} \Gamma_{\gamma}^{\ast}(Te) C_{\gamma} = -  \Gamma_{\gamma}(Te)$, Eq.~(\ref{eq:chig}) revisits the previous result~\cite{kobayashi224504}. 
For two MFs, $\chi^{\Omega_{-,h}}(g)$ yields the same results as those in Table~\ref{classification_type3}. The short representations of two MFs for the type-(I) and (III) groups are summarized in Tables~\ref{tab:short1} and~\ref{tab:short3}.

We comment on the relationship between $\Gamma_{\gamma}(Th)$ and the 1D winding number $W[h]^g_j$~\cite{yamazaki073701}. The bulk-boundary correspondence yields~\cite{Sato224511, Izumida195442, xiong17} 
\begin{align}
   \text{tr}(\Gamma_{\gamma,j}(Th)) = W[h]^g_j,
\end{align}
where $\Gamma_{\gamma,j}(Th)$ is $\Gamma_{\gamma}(Th)$  defined in the eigenspace of $g$.
When $g$ and $h$ are order-two crystalline operators, such as two-fold rotation and mirror-reflection operators, the number of MFs $N_{2,\text{W}}$ is directly related to $W[h]^g_j$ as~\cite{yamazaki073701}:
\begin{align}
    N_{2,\text{W}} &= -i \text{tr}(\Gamma_{\gamma}(Th)\tilde{D}_{\gamma}(g)) \nonumber \\
    &= \text{tr}(\Gamma_{\gamma,1}(Th)) - \text{tr}(\Gamma_{\gamma,2}(Th)) \nonumber \\
    &= W[h]^g_1 - W[h]^g_2.
\end{align}

 We extend the relationship to the cases with $T4_z$ and $T6_z$ symmetries. In $T4_z$ symmetric systems, the 1D winding number is defined by
\begin{align}
    W[4_z]^{2_z}_{j} = \frac{i}{4\pi} \int_{-\pi}^\pi dk \tr\left\{\Gamma_j(T4_z) H_j^{-1}(k) \pdv{H_j(k)}{k}\right\},
\end{align}
with
\begin{align}
    &\Gamma(T4_z)  \to \Gamma_1(T4_z) \oplus \Gamma_2(T4_z),  \\
    &H(k) \to H_1(k) \oplus H_2(k), 
\end{align}
where $[\tilde{D}(2_z),H(k)]=[\tilde{D}(2_z),\Gamma(T4_z)]=0$. We choose the phase of $\Gamma_{\gamma}(T4_z)$ as $\Gamma^4(T4_z) = 1$, implying that  $\Gamma_1^2(T4_z) = e^{-2i\phi_{4,1}}$ and $\Gamma_2^2(T4_z) = e^{-2i\phi_{4,2}}$ with $4\phi_{4,j}= 0 \mod 2 \pi$, and $e^{i \phi_{4,j}}W[4_z]^{2_z}_{j} \in \mathbb{Z}$.
By using the 1D winding numbers, $\text{tr} (\Gamma_{\gamma}(T4_z)\tilde{D}(2_z)) $ is described as
\begin{align}
    \text{tr} &(\Gamma_{\gamma}(T4_z)\tilde{D}_{\gamma}(2_z)) \nonumber \\
     &= i(W[4_z]^{2_z}_1 - W[4_z]^{2_z}_2) \nonumber \\
    &=i(e^{-i \phi_{4,1}} + e^{-i \phi_{4,2}}) e^{i\phi_{4,1}}W[4_z]^{2_z}_1
\end{align}
where we have used $e^{i \phi_{4,1}} W[4_z]^{2_z}_{1} = -e^{i \phi_{4,2}} W[4_z]^{2_z}_{2}$ due to $\tilde{D}(4_zT)$. Thus, when the number of MFs is defined as 
\begin{align}
N_{4,\text{W}} &= e^{i \phi_{4,1}} W[4_z]^{2_z}_{1}  - e^{i \phi_{4,2}} W[4_z]^{2_z}_{2}\nonumber \\
                &= 2 e^{i \phi_{4,1}} W[4_z]^{2_z}_{1} ,
\end{align}
we obtain 
\begin{align}
 |N_{4,\text{W}}| = 2 \frac{|\text{tr} (\Gamma_{\gamma}(T4_z)\tilde{D}_{\gamma}(2_z))|}{ |1+e^{i (\phi_{4,1} - \phi_{4,2})}| },
\end{align}
where $e^{i (\phi_{4,1} - \phi_{4,2}) } \neq -1$. 

Similarly, in $T6_z$ symmetric systems, the 1D winding number is defined by
\begin{align}
    W[6_z]^{3_z}_{j} = \frac{i}{4\pi} \int_{-\pi}^\pi dk \tr\left\{\Gamma_j(T6_z) H_j^{-1}(k) \pdv{H_j(k)}{k}\right\},
\end{align}
with
\begin{align}
    &\Gamma(T6_z)  \to \Gamma_1(T6_z) \oplus \Gamma_2(T6_z) \oplus \Gamma_3(T6_z),  \\
    &H(k) \to H_1(k) \oplus H_2(k) \oplus H_3(k), 
\end{align}
where $[\tilde{D}(3_z),H(k)]=[\tilde{D}(3_z),\Gamma(T6_z)]=0$ and $\Gamma^6(T6_z) = 1$, i.e., $6 \phi_{6,j} = 0 \mod 2\pi$ for the phase of $\Gamma_j(T6_z)$. The relationship among $\Gamma_{\gamma}(T6_z)$, $W[6_z]^{3_z}_{j}$ and the number of MFs $N_{6,\text{W}}$ is given by
\begin{align}
    &\text{tr}(\Gamma_{\gamma}(T6_z)\tilde{D}_{\gamma}(3_z)) \nonumber \\
    &\qquad = e^{-i\frac{\pi}{3} } W[6_z]^{3_z}_1 +e^{-i\frac{5\pi}{3} } W[6_z]^{3_z}_2  - W[6_z]^{3_z}_3, \\
    &N_{6,\text{W}} = 2 e^{i \phi_{6,1}} W[6_z]^{3_z}_{1} + e^{i \phi_{6,3}} W[6_z]^{3_z}_{3},
\end{align}
where we have used $e^{i \phi_{6,1}} W[6_z]^{3_z}_{1} = -e^{i \phi_{6,2}} W[6_z]^{3_z}_{2}$ due to $\tilde{D}(6_zT)$.
To proceed the calculation, we consider two cases: when $ W[6_z]^{3_z}_3=0$, the relation is given by
\begin{align}
    |N_{6,\text{W}}| = 2 \frac{|\text{tr} (\Gamma_{\gamma}(T6_z)\tilde{D}_{\gamma}(3_z))|}{ |1+e^{-i\frac{\pi}{3}}e^{i (\phi_{6,1} - \phi_{6,2}) } |},
\end{align}
where $e^{i (\phi_{6,1} - \phi_{6,3})} \neq e^{-\frac{2\pi}{3}} $, while, when $ W[6_z]^{3_z}_1=W[6_z]^{3_z}_2=0$, it becomes
\begin{align}
    |N_{6,\text{W}}| =|\text{tr} (\Gamma_{\gamma}(T6_z)\tilde{D}_{\gamma}(3_z))|.
\end{align}

\section{Magnetic tetrahexacontapole response}
\label{sec:tetrahexacontapole}
We consider a tight-binding model with the magnetic point group $6/mm'm'$, which is built on the triangular lattice with $p_x$, $p_y$, and $p_z$ orbitals and two sublattice degrees of freedom on each site. In the BdG Hamiltonian (\ref{BdG22}), the normal-state Hamiltonian and gap function are given by 
\begin{align}
\epsilon(\bm{k}) = \epsilon_0(\bm{k}) + \epsilon_{\text{soc}}(\bm{k}) +M_z s_z ,
\end{align}
with
\begin{align}
\nonumber
\epsilon_0(\bm{k}) &= m_0 + m_1 \lambda_8 + t_z \cos k_z + t'_z \sin k_z \sigma_y \\
&\quad + t_{xy} \bigg\{\cos k_x + 2\cos \left(\frac{k_x}{2}\right)\cos \left(\frac{\sqrt{3}k_y}{2}\right)  \bigg\}, \\ \nonumber
\epsilon_{\text{soc}}(\bm{k}) &= \alpha \cos k_z (\lambda_5 s_y -\lambda_7 s_x) \\
&\quad + \alpha'\sin k_z \sigma_y (\lambda_5 s_y -\lambda_7 s_x),
\end{align}
and $\Delta(\bm{k}) = \tilde{\Delta}(\bm{k}) i s_y$ with
\begin{align}
\nonumber
\tilde{\Delta}(\bm{k}) = &\Delta_{B_{2u}} \sin k_z (\lambda_1 s_y-\lambda_3 s_x) \\ 
&- i\Delta_{B_{1u}} \sin k_z (\lambda_1 s_x+\lambda_3 s_y).
\end{align}
Here, $\lambda_i$ ($i = 1, \ldots, 8$) are the Gell-Mann matrices acting on the ($p_x$, $p_y$, $p_z$) orbitals, and $\sigma_i$ and $s_i$ are the Pauli matrices in the sublattice and spin spaces. 
In the normal-state Hamiltonian, $m_0$ and $m_1$ represent onsite potentials, $t_z$, $t_z'$, and $t_{xy}$ hopping parameters, and $\alpha$ and $\alpha'$ spin-orbital interactions. The gap function belongs to the $B_u$ IR of $6/mm'm'$ ($H_0=6/m$), where the $\Delta_{B_{2u}}$ and $\Delta_{B_{1u}}$ terms are basis functions that follows $B_{2u}$ and $B_{1u}$ IRs of $6/mmm$.  

The symmetry operations of $6/mm'm'$ are given by
\begin{align}
D(6_z) &= \sigma_0 R_z\left( \frac{2\pi}{6} \right) \exp\left( -i s_z \frac{\pi}{6} \right), \\
D(I) &= -\sigma_x, \\
D(T2_y) &= -\sigma_x R_y(\pi)K, \\
D(T2_x) &= \sigma_x R_x(\pi)(-i s_z)K,
\end{align}
where $R_i(\theta)$ is a $3 \times 3$ rotation matrix in the basis ($p_x$, $p_y$, $p_z$) and represents the $\theta$ rotation about the $i$ axis, and $K$ is the complex conjugation. 

On the $(xy)$ surface, the 2D magnetic point group compatible with the surface is $M^{\text{(III)}}_0 = 6m'm'$ and the IR of the gap function is the $B$ IR of $6m'm'$ ($H_0=6$). In this case, we have two 1D winding numbers $W[m_{yz}]$ and $W[m_{zx}]$, which are defined by
\begin{align}
    W[h] &= \frac{i}{4\pi} \int_{-\pi}^\pi dk \tr\left\{\Gamma(Th) H^{-1}(k) \pdv{H(k)}{k}\right\}, \nonumber \\
    &= \int^{\pi}_{-\pi} \frac{dk_z}{2\pi} \frac{\partial}{\partial {k_{z}}} \text{arg} \{\text{det}[q(k_z)]\},
\label{eq:winding-6mm}
\end{align}
where $h=m_{yz}$ and $m_{zx}$. The chiral operators are given by 
\begin{align}
\Gamma(Tm_{yz}) = R_x(\pi)s_z\tau_y, \  \Gamma(Tm_{zx}) = R_y(\pi)\tau_y.
\end{align}
where $\tau_i$ ($i=x,y,z$) are the Pauli matrices in the Nambu space. The factor $q(k_z)$ is defined as in Eq.~(\ref{eq:def_q}).

\begin{figure}[t]
\vspace{5mm}
\centering
    \includegraphics[scale=0.37]{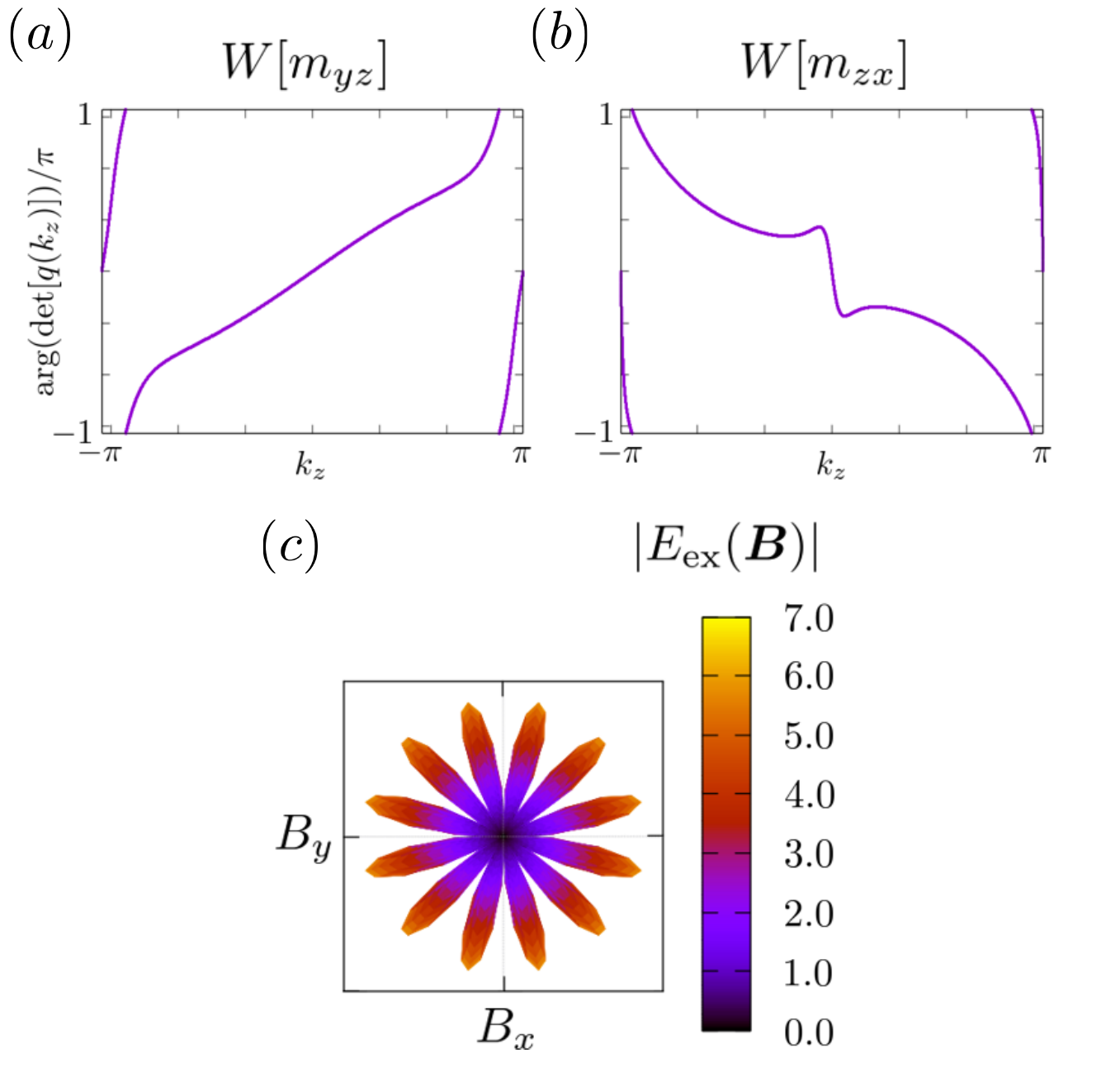}
    \caption{ In (a) and (b), we show the change in $q(k_z)$ in Eq.~(\ref{eq:winding-6mm}) as a function of $k_z$ at $k_x=k_y=0$: (a) $W[m_{yz}]$ and (b) $W[m_{zx}]$.  In (c), we illustrate the polar plot of the energy gap at $k_x=k_y=0$ under the Zeeman magnetic field. The color shows the normalized magnitude of the energy gap $E_{\text{ex}}(\bm{B})$. }
    \label{Fig: tetrahexacontapole}
\end{figure}

We numerically demonstrate the magnetic response of MFs. We choose the parameters as $m_0=-1.0,t_z=3.0,t'_z=0.1,t_{xy}=0.3,m_1=0.1,\alpha=4.0,\alpha'=1.0,M_z=3.0,\Delta_{B_{1u}}=0.3,\Delta_{B_{2u}}=0.4$. In these parameters, two MFs emerge at $k_x=k_y=0$. Figures~\ref{Fig: tetrahexacontapole} (a) and (b) show the phase change of $q(k_z)$ in Eq.~(\ref{eq:winding-6mm}), which indicates $|W[m_{yz}]| = |W[m_{zx}]| =2$. Therefore, the MFs are characterized by the two 1D winding numbers, which may exhibit the magnetic tetrahexacontapole response, as indicated by Table~\ref{classification_type3}.  We add the Zeeman magnetic term as in Eq.~(\ref{surface-magnetic-fields}) with $B=2.0$. Figure~\ref{Fig: tetrahexacontapole} (c) displays the magnetic tetrahexacontapole response, where the energy gap of the MFs is described as
\begin{align}
  E_{\text{ex}}(\bm{B}) \sim (B_x^3-3 B_x B^2_y)(B_y^3-3 B_y B^2_x).
\end{align}
The result is consistent with our classification.

\section{Strain response of MFs}
\label{sec:strain-response}

\begin{figure}[t]
\vspace{5mm}
\centering
    \includegraphics[scale=0.37]{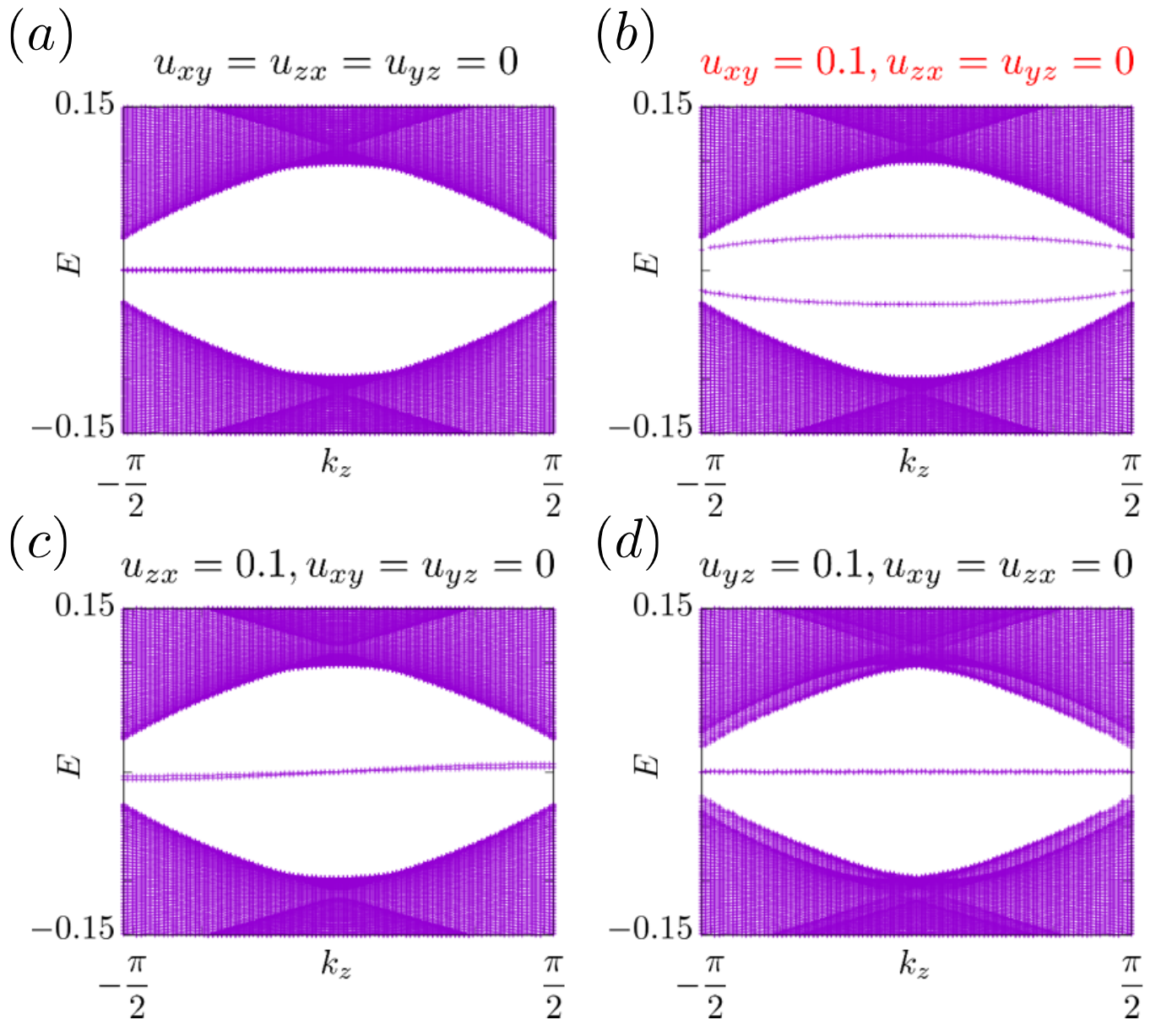}
    \caption{The surface energy spectrum of the BdG Hamiltonian~(\ref{eq:bdg_open}) with the strain (\ref{surface-strain-fields}) in the $(zx)$ surface at $k_x=0$, where the parameters are chosen to be (a) $(u_{xy},u_{zx},u_{yz}) = (0,0,0)$, (b) $(0.1,0,0)$, (c) $(0,0.1,0)$, and (d) $(0,0,0.1)$, respectively.}
    \label{Fig: energy-electric-response}
\end{figure}

As shown in Tables~\ref{classification_type1} and \ref{classification_type3}, MFs can couple to strain, which is a time-reversal-even perturbation. Thus, strain response of two MFs indicates the breaking of TRS.
To verify the strain response of MFs, we consider the model of UTe$_2$ in Sec. \ref{subsec: UTe2}, which exhibits the magnetic quadrupole response. We add the strain term:
\begin{align}
\epsilon_{\text{strain}}(\bm{k})= (u_{xy}s_y \sigma_z+ u_{zx}s_z \sigma_z + u_{yz}\sigma_y) j(\bm{k})
\label{surface-strain-fields}
\end{align} 
to Eq.~(\ref{normalpart1}) and numerically diagonalize the BdG Hamiltonian as in Eq.~(\ref{eq:bdg_open}). Figures~\ref{Fig: energy-electric-response} (a)--(d) show the energy spectra of the BdG Hamiltonian under the strain (\ref{surface-strain-fields}), where we choose the parameters $(u_{xy}, u_{zx}, u_{yz})$ as (a) $(0,0,0)$, (b) (0.1,0,0), (c) (0,0.1,0), and (d) (0,0,0.1). The results indicate that the coupling between MFs and the $u_{xy}$ term significantly changes the energy spectrum and opens a gap at $k_x=k_z=0$. The energy gap of the MFs under the strain is described as
\begin{align}
    E_{\text{ex}}(\bar{u}) \sim u_{xy},
\end{align}
which is consistent with Table~\ref{classification_type3}.

\newpage
\bibliography{ref} 
\end{document}